\documentclass[11pt,draft]{article}
 \usepackage{amsfonts}
 \usepackage{amssymb}
 \def\bsh{\backslash}

 \newfont{\bbbold}{msbm10}

 \def\bbR{\mbox{\bbbold R}}

 \def\cM{{\cal M}}

 \def\cU{{\cal U}}
 \def\cV{{\cal V}}

 \newfont{\goth}{eufm10 scaled \magstep1}

 \def\gh{\mbox{\goth h}}

 \def\a{\alpha}
 \def\b{\beta}
 \def\c{\gamma}
 
 \def\d{\delta}
 \def\dh{{\hat{\delta}}}
 \def\e{\epsilon}
 
 \def\ah{{\hat\alpha}}
 \def\bh{{\hat\beta}}
 \def\gh{{\hat\gamma}}
 \def\dh{{\hat\delta}}
 \def\hp{\hat p}
 \def\hht{\hat \theta}
 \def\hd{\hat d}
 \def\hl{{\hat \lambda}}
 \def\hw{\hat w}
 \def\e{\epsilon}
 
 \def\g{\gamma}
 \def\h{\eta}
 \def\k{\kappa}
 \def\l{\lambda}\def\L{\Lambda}

 \def\p{\partial}
 \def\P{\Pi}
 
 \def\s{\sigma}\def\S{\Sigma}
 \def\t{\theta}
 \def\th{\theta}
 
 \def\be{\begin{equation}}\def\ee{\end{equation}}
 \def\bea{\begin{eqnarray}}\def\eea{\end{eqnarray}}
 \def\ba{\begin{array}}\def\ea{\end{array}}
 
 \def\o{\omega}\def\O{\Omega}

 \def\xz{\times}
 \def\half{{1\over 2}}
 \def\pa{\partial}

 \def\nab{\nabla}
 \def\tD{\tilde{D}}\def\tF{\tilde{F}}
 \def\tT{\tilde{T}}\def\tR{\tilde{R}}\def\tO{\tilde{\Omega}}
 \def\hO{\hat{\Omega}}

 \def\3dt{\dot{3}}

 \let\la=\label

 \let\bm=\bibitem{}

 \def\nn{\nonumber}
 \def\bd{\begin{document}}
 \def\ed{\end{document}}
 \def\bea{\begin{eqnarray}}\def\barr{\begin{array}}\def\earr{\end{array}}
 \def\eea{\end{eqnarray}}
 \def\ft#1#2{{\textstyle{{\scriptstyle #1}\over {\scriptstyle #2}}}}
 \def\fft#1#2{{#1 \over #2}}
 \newcommand{\eq}[1]{(\ref{#1})}
 \def\eqs#1#2{(\ref{#1}-\ref{#2})}
 \def\det{{\rm det\,}}
 \def\tr{{\rm tr}}\def\Tr{{\rm Tr}}

\begin{document}
 %%%%%%%%%%%%%%%%%%%%%%%%%%%%%%%%%%%%%%%%%%%%%%%%%%%%%%%%%%%%%%%%%%%%%%%%%%%%%

\thispagestyle{empty}

 \hfill{IFT-P.072/2001}

 \hfill{KCL-TH-01-49}

 \hfill{hep-th/0112160}

 \hfill{\today}

 \vspace{20pt}

 \begin{center}
 {\Large{\bf Ten-Dimensional Supergravity Constraints from the
Pure Spinor Formalism for the Superstring}}
 \vspace{30pt}

  {Nathan Berkovits}
  \vskip .2cm {Instituto 
de F\'{\i}sica Te\'orica, Universidade Estadual Paulista}
  \linebreak {S\~ao Paulo, Brasil} \vspace {15pt}
  \vskip .5cm
   {and}
  \vskip .5cm
  {Paul Howe} \vskip .2cm {Department of Mathematics}\linebreak
  {King's College, London}

  \vspace{60pt}
 
  \end{center}

 {\bf Abstract}
 
\par
 It has recently been shown that the ten-dimensional superstring can
 be quantized using the BRST operator $Q=\oint\l^\alpha d_\alpha$ where
 $\l^\alpha$ is a pure spinor satisfying $\l \c^m \l=0$ and $d_\alpha$
 is the fermionic supersymmetric derivative. In this paper, 
 the pure spinor version of superstring theory is
 formulated in a curved supergravity background and it is shown that 
 nilpotency and holomorphicity of the pure spinor BRST operator
 imply the on-shell superspace constraints of the supergravity
 background.
 This is shown to lowest order in $\alpha'$ for the
 heterotic and Type II superstrings, thus providing a 
 compact pure spinor version of the ten-dimensional superspace
 constraints for $N=1$, Type IIA and Type IIB supergravities. Since
 quantization is straightforward using the pure spinor version of
 the superstring, it is expected that these methods can
 also be used to compute higher-order $\alpha'$ corrections to the
 ten-dimensional superspace constraints.
 
 {\vfill\leftline{}\vfill \vskip  10pt
 
  \baselineskip=15pt \pagebreak \setcounter{page}{1}
 
 \section{Introduction}
 
For many purposes, superstring theory is most conveniently expressed as
 an effective field theory of its massless modes consisting of
 supergravity theory together with corrections arising order by
 order in $\a'$. In principle these higher order corrections can be
 obtained by computing scattering amplitudes or by demanding
 consistency of the superstring sigma model in a curved
 background. However, neither of these procedures is easy to carry
 out in superstring theory in a way in which spacetime supersymmetry
 is guaranteed. In the Ramond-Neveu-Schwarz (RNS) formalism, 
it is difficult to introduce fermionic or
Ramond-Ramond background
 fields, while the Green-Schwarz (GS) formalism
 ensures spacetime supersymmetry but is difficult to
 quantise. Although the hybrid formalism for the superstring
can be used to compute $\a'$ corrections in a manner which manifestly
preserves
D=2 \cite{twoh}, D=4 \cite{fourh}, 
or D=6 \cite{sixh} super-Poincar\'e
covariance,
one needs a D=10 covariant formalism if one wants to describe the
superstring in arbitrary supergravity backgrounds.
 
 In this situation, one might try to study the
 constraints that ten-dimensional supersymmetry imposes on higher-order
 contributions to the effective action. One difficulty here is that,
 with the exception of the supergravity sector of the heterotic string,
 it is not known how to construct any superspace actions due to the
 absence of any known sets of auxiliary fields. Even in  the
 heterotic case, the auxiliary fields are rather complicated \cite{hnvp} and it
 is not clear how to construct higher order actions which correspond
 to superstring 
corrections, although the $R^4$ invariant was discussed 
from this point of view 
in \cite{nt}. It seems that additional input apart from
 supersymmetry is required. On the other hand, it has been possible
 to obtain information about some particular terms, for example in
 the work of 
 \cite{vw,s,gg,aetal,rt,bv,gs}. Other approaches to the problem have involved
 supersymmetrisation of 
bosonic sigma model terms \cite{gvdvz} for the heterotic
 string \cite{bdr,drsw}, 
and studying corrections to heterotic superspace constraints
 directly 
\cite{betal,getal}, which has at least been successful in incorporating
 anomaly terms. This work has been reviewed in \cite{pvhw} where string 
results were used to partially construct $R^4$ corrections in M-theory.
Other recent approaches to supersymmetrizing the $R^4$ term in M-theory are
described in \cite{ceder} and \cite{gatesm}.

 The fact that one is forced to look at the equations of motion
 rather than Lagrangians suggests that a way forward might be 
 to understand the geometry behind these equations. Many years ago,
 Witten showed how the N=1 D=10 superspace Yang-Mills equations
 can be understood in terms of integrability along light-like lines
 and how this is related to $\k$-symmetry of the superparticle
 action \cite{w,st}. This sort of analysis was subsequently carried out for the
 heterotic \cite{chau}
 and IIB strings \cite{ghnt}, and reinterpreted in terms of
 light-like 
integrability in loop superspace in \cite{bhpss,bds}, at least for the
 heterotic case.
 
 In some related work, one of the present authors showed that
 light-like integrability could be replaced by integrability along
 pure spinor lines, and that this can also be employed in eleven
 dimensions in the context of the supermembrane \cite{h}. A virtue of this
 approach is that it is simpler than light-like integrability, but,
 at the time, it wasn't entirely clear how it was related to
 particle or string actions. More recently, the other  author
 has shown that ten-dimensional
superparticles and superstrings can be effectively
 quantised using pure spinor variables \cite{superp,
me}. These pure spinor variables
can be interpreted as bosonic ghosts for a fermionic symmetry,
 although it is not currently fully understood
 how this can be implemented in a worldsheet reparameterization
invariant fashion. Nevertheless,
 the final gauge-fixed action does have manifest spacetime
 supersymmetry and correctly fixes the central charge to be zero.
 Unlike the GS formalism,
 however, the pure spinor formalism has the tremendous advantage
 that it can be quantised straightforwardly since the action is
 free in a flat background.

 In this pure spinor formalism \cite{me}, the left-moving
BRST operator for the heterotic
superstring is
 \be
 Q=\oint \lambda^\a d_\a 
\label{BRSTh}
\ee 
where $\lambda^\a$ is a bosonic
 pure spinor variable satisfying\footnote{We will use the notation
 where $\g^m_{\a\b}$ and $\g^{m~\a\b}$ are $16\times 16$ symmetric
 matrices which form the off-diagonal blocks of the $32\times 32$
 ten-dimensional $\Gamma$-matrices in the Weyl representation.}
 \be
 \lambda^\a \g_{\a\b}^m \lambda^\b=0 
\label{pure} 
\ee 
for $m=0$ to 9,
 and $d_\a$ is the worldsheet variable corresponding to the
 N=1 D=10 spacetime supersymmetric derivative.
In a flat background, $\l^\a$ and
 $d_\a$ are holomorphic and $d_\a$ satisfies the OPE $d_\a(y)
 d_\b(z) \to -i \a'(y-z)^{-1}\g^m_{\a\b} \Pi_m$ where $\Pi_m = \p x_m
 +{i\over 2}\th\g_m\p\th$ is the supersymmetric momentum. So
 $\l\g^m\l=0$ implies that $Q$ is nilpotent.
 A natural conjecture is that in a curved supergravity/super-Yang-Mills
 background for the heterotic superstring,
 nilpotence and holomorphicity of $\l^\a d_\a$ implies the
 superspace equations of motion for the background superfields.
 
Similarly, in the pure spinor formalism for the Type II superstring,
the left and right-moving BRST operators are 
\footnote{Throughout this paper, we will use spinor notation simultaneously
for the Type IIA and Type IIB superstring by imposing that $\a$ and $\ah$
denote D=10 spinors of opposite chirality for the IIA superstring and
denote spinors of the same chirality for the IIB superstring.}
 \be
 Q=\oint \lambda^\a d_\a,\quad
 \bar Q=\oint \hat\lambda^\ah \hat d_\ah,
\label{BRSTII}
 \ee 
where $\l^\a$ and $\hl^\ah$ are independent pure spinor variables
satisfying
 \be
 \lambda^\a \g_{\a\b}^m \lambda^\b=0, \quad 
 \hat\lambda^\ah \g_{\ah\bh}^m \hat\lambda^\bh=0, 
\label{pureeII} 
\ee
for $m=0$ to 9, and $d_\a$ and $\hat d_\ah$ are worldsheet variables
corresponding to the N=2 D=10 spacetime supersymmetric derivatives.
In a flat background, $\l^\a d_\a$ is holomorphic and nilpotent whereas
$\hl^\ah \hat d_\ah$ is antiholomorphic and nilpotent. So it is natural
to conjecture that in a curved N=2 D=10 supergravity background for the
Type II superstring, the superspace equations of motion for the background
are implied
by the condition that these properties of $\l^\a d_\a$ and 
$\hl^\ah \hat d_\ah$ are preserved. 

 In this paper, we shall verify the above conjectures to lowest order
 in $\a'$ for the heterotic and Type II superstrings in $N=1$ and
 $N=2$ supergravity backgrounds. This verification will lead to new
 pure spinor versions of the superspace constraints for
 ten-dimensional $N=1$, Type IIA and Type IIB supergravity. 
These have the property that they are remarkably compact 
and may be useful for studying other aspects
 of ten-dimensional supersymmetric theories such as harmonic
 superspace.
 Furthermore, since the superstring action is quantizable, 
this conjecture can be
 used in principle to compute the superspace equations of motion
 to arbitrary order in $\alpha'$.
 
For the N=1 D=10 supergravity/super-Yang-Mills background of the 
heterotic superstring, nilpotence of $\l^\a d_\a$ will imply
\be
\l^\a \l^\b T_{\a\b}{}^C= \l^\a\l^\b H_{\a\b C }= \l^\a\l^\b F_{\a\b}^I=0
\label{pureintro}
\ee
where $T_{AB}{}^C$, $H_{ABC}$ and $F_{AB}^I$ are the superspace
torsion, three-form field strength and super-Yang-Mills field strength. 
These equations are 
identical to those derived from pure spinor integrability in \cite{h}.
Since (\ref{pureintro}) must be satisfied for an arbitrary pure spinor
$\l^\a$ satisfying (\ref{pure}), (\ref{pureintro}) implies that 
\be
(\g_{mnpqr})^{\a\b} T_{\a\b}{}^C=(\g_{mnpqr})^{\a\b} H_{\a\b C }
=(\g_{mnpqr})^{\a\b}  F_{\a\b}^I=0
\label{pureintrotwo}
\ee
for any self-dual five-form direction $mnpqr$. Up to conventional constraints
(which will be implied by holomorphicity of $\l^\a d_\a$), the
constraints of (\ref{pureintrotwo}) will be shown to imply the standard N=1
supergravity/super-Yang-Mills equations of motion.

For the N=2 D=10 supergravity background of the 
Type II superstring, nilpotence of $\l^\a d_\a$ and $\hl^\ah \hat d_\ah$ 
will imply
\be
\l^\a \l^\b T_{\a\b}{}^C= 
\hl^\ah \hl^\bh T_{\ah\bh}{}^C= 
\l^\a \hl^\bh T_{\a\bh}{}^C= 0,
\label{pureIIintro}
\ee
$$\l^\a\l^\b H_{\a\b C }= 
\hl^\ah\hl^\bh H_{\ah\bh C }= 
\l^\a\hl^\bh H_{\a\bh C }= 0.$$
Since 
$\l^\a$ and $\hl^\ah$ are arbitrary pure spinors
satisfying (\ref{pureeII}), (\ref{pureIIintro}) implies that 
\be
(\g_{mnpqr})^{\a\b} T_{\a\b}{}^C= 
(\g_{mnpqr})^{\ah\bh} T_{\ah\bh}{}^C= 
 T_{\a\bh}{}^C= 0,
\label{pureIIintrotwo}
\ee
$$(\g_{mnpqr})^{\a\b} H_{\a\b C }= 
(\g_{mnpqr})^{\ah\bh} H_{\ah\bh C }= 
H_{\a\bh C }= 0$$
for any self-dual five-form direction $mnpqr$.
Up to conventional constraints (which will be implied by holomorphicity
and antiholomorphicity of $\l^\a d_\a$ and $\hl^\ah \hat d_\ah$),
the constraints of (\ref{pureIIintrotwo}) will be shown to imply the standard 
Type II supergravity equations of motion. \footnote{As will be explained in
section 3, the superspace torsion $T_{AB}{}^C$ appearing in 
(\ref{pureIIintro})
and (\ref{pureIIintrotwo}) is not the usual one since some
of its components depend on a ``left-moving'' spin connection and
some of its components depend on a ``right-moving'' spin connection.}

 In section 2 of this paper we shall use the heterotic superstring
 sigma model to derive a pure spinor version of the $N=1$
 supergravity/super-Yang-Mills constraints, and in section 3 we
 shall use the Type II superstring sigma model to derive a pure
 spinor version of the Type IIA and Type IIB supergravity
 constraints. In section 4  the pure spinor description of Type IIB
 supergravity will be shown to agree with the
 standard Howe-West (HW)
superspace description of \cite{hw}. 
 In section 5 we shall briefly
 discuss the the procedure for extending to higher order in $\a'$
 these computations of the ten-dimensional superspace constraints.

\section{Heterotic Superstring Sigma Model}

In this section, the pure spinor version of the heterotic superstring
will be reviewed in flat and curved backgrounds. Nilpotence and
holomorphicity of $\l^\a d_\a$ will then be shown to imply the
superspace equations of motion for the 
supergravity/super-Yang-Mills background.

\subsection{Heterotic superstring in a flat background}

In the pure spinor version of the heterotic superstring, the worldsheet
variables consist of the $N=1$ $D=10$
superspace variables $(x^m,\th^\a,p_\a)$ for $m=0$ to 9 and
$\a=1$ to 16 where $p_\a$ is the conjugate momentum to $\th^\a$, as well
as
the left-moving pure spinor ghost variable $\l^\a$ and its conjugate momentum
$w_\a$,
the $E_8\times E_8$ or $SO(32)$ right-moving currents $\bar J^I$,
and $(\bar b,\bar c)$ right-moving Virasoro ghosts.
Because $\l^\a$ is defined to satisfy (\ref{pure}), it has only
eleven independent degrees of freedom and its conjugate
momentum $w_\a$ is only defined up to the gauge transformation
$\d w_\a=\Lambda_m (\g^m\l)_\a$ for any $\Lambda^m$. This gauge
transformation can be used to eliminate five components of $w_\a$,
so both $\l^\a$ and $w_\a$ have 
eleven independent components.

The action and stress-tensor in a flat background is
\be
S={1\over{2\pi \a'}}\int d^2 z (\half \pa x^m \bar\pa x_m + p_\a \bar\pa\th^\a 
+ \bar b \pa\bar c) + S_\l + S_J,
\label{flataction}
\ee
\be
T = {1\over \a'}(-\half \pa x^m \pa x_m -p_\a \pa\t^\a) + T_\l,\quad
\bar T = {1\over \a'}(-\half \bar\pa x^m \bar\pa x_m  -\bar b\bar\pa\bar c
-\bar\pa(\bar b\bar c)) + T_J,
\ee
where $S_\l$ and $S_J$ are the actions for $\l^\a$ and $J^I$,
and $T_\l$ and $T_J$ are the $c=22$ and $c=16$ stress tensors for
$\l^\a$ and $J^I$.
As described in \cite{me}, one can write explicit expression for $S_\l$ and
$T_\l$ by solving the constraint of (\ref{pure})
in terms of eleven
chiral bosons $(\g,u_{ab})$ and their conjugate momenta $(\b,v^{ab})$
where $a=1$ to 5 and $u_{ab}=-u_{ba}$. 
However, these explicit expressions will not be necessary here. We will
only need to know that $S_\l$ is defined such that $\l^\a$ has no singular
OPE with itself and Lorentz currents $N^{mn}$ can be
constructed out of $\l^\a$ and its conjugate momentum $w_\a$
as $N^{mn}={1\over{2\a'}} \l\g^{mn}w$ which satisfy the
OPE's
\be
N^{mn}(y) \l^\a (z) \to
\half(\g^{mn})^\a{}_\b {{\l^\b(z)}\over{y-z}}
\label{OPEl}
\ee
\be
N^{kl}(y) N^{mn}(z) \to
{{\eta^{m[l} N^{k]n}(z) -
\eta^{n[l} N^{k]m}(z) }\over {y-z}} - 3
{{\eta^{kn} \eta^{lm} -
\eta^{km} \eta^{ln}}\over{(y-z)^2}}  .
\label{OPEN}
\ee
Similarly, the explicit expression for $S_J$ will not be necessary
and we will only need to know the OPE
\be
J^I(y) J^K(z) \to {{\d^{IK}}\over{(y-z)^2}} + f^{IK}_L {{J^L(z)}\over
{(y-z)}}
\ee
where $f^{IK}_L$ are the $E_8\times E_8$ or $SO(32)$ structure constants.

Physical states of the superstring are defined as vertex operators
in the cohomology of the left and right-moving BRST operators
\be
Q= \oint \l^\a d_\a,\quad
\bar Q = \oint (\bar c \bar T + \bar c \bar\pa\bar c \bar b)
\ee
where 
\be
d_\a = p_\a -{i\over 2}\g_{\a\b}^m \t^\b \p x_m +{1\over 8}
\g_{\a\b}^m \g_{m~\g\d}\t^\b\t^\g\p\t^\d {\rm ~~~~ and ~~~~}
\Pi^m = \p x^m +{i\over 2}\t\g^m\p\t
\ee
are spacetime supersymmetric and satisfy the OPE's \cite{siegel}
\be
d_\a(y) d_\b(z) \to -i\a'(y-z)^{-1} \g_{\a\b}^m \Pi_m(z),\quad
d_\a(y) \Pi^m(z) \to  i\a' (y-z)^{-1} \g_{\a\b}^m \p\t^\b(z).
\label{OPEd}
\ee

To construct the sigma model for the heterotic superstring,
it will be useful to know the integrated form of the massless
supergravity and super-Yang-Mills vertex operators, $V_{SG}$ and
$V_{sYM}$, which are
\be
V_{SG} = \int d^2 z (\pa\th^\a A_{\a m}(x,\th) + 
\Pi^n A_{nm}(x,\th) + d_\a E^\a_m (x,\th) +\half N_{np} \Omega_m^{np}(x,\th))
\bar\p x^m,
\label{vsg}
\ee
\be
V_{sYM}=\int d^2 z (
\pa\th^\a A_{\a I}(x,\th) + 
\Pi^n A_{n I}(x,\th) + d_\a W^\a_I (x,\th) +\half N_{np} U_I^{np}(x,\th))
\bar J^I,
\label{vsym}
\ee
where $N^{np}$ are the Lorentz currents for the pure spinor.
Note that the first two terms in $V_{SG}$ and $V_{sYM}$ are the
same as in the Green-Schwarz heterotic superstring vertex operators,
but the third and fourth terms are needed for the vertex operators
to be BRST invariant. These two vertex operators can be obtained
by taking the ``product'' of a massless
open superstring vertex operator,
\be
V_{open}=\int dz (
\pa\th^\a A_{\a}(x,\th) + 
\Pi^n A_{n }(x,\th) + d_\a W^\a (x,\th) +\half N_{np} U^{np}(x,\th)),
\label{open}
\ee
with either $\int d\bar z \bar\p x^m$ or $\int d\bar z \bar J^I$.

Using the fact that $\l^\a\l^\b$ is proportional to $\g_{mnpqr}^{\a\b}
(\l\g^{mnpqr}\l)$ and
the OPE's of (\ref{OPEl}) and (\ref{OPEd}), one can check
that $QV_{SG} = \bar Q V_{SG}=0$ implies that 
\be
\g_{npqrs}^{\a\b} D_\a A_{\b m} = 0,\quad
\p^m(\pa_m A_{\b n} - \pa_n A_{\b m})=0,
\label{sugl}
\ee
\be
A_{nm}=-{i\over 8} D_\a \g_n^{\a\b} A_{\b m},~
E^\b_m = -{i\over{10}}\g^{n\a\b} (  D_\a A_{nm} -\pa_n A_{\a m}),~
\Omega_m^{np} =
{1\over 8}D_\a (\g^{np})^\a{}_\b E^\b_m = \pa_{[n} A_{p]m}
\label{sugc}
\ee
where $D_\a = {\p\over{\p\t^\a}} +{i\over 2} \g^m_{\a\b}\t^\b\p_m$
is the N=1 D=10 supersymmetric derivative.
Similarly, 
$QV_{sYM} = \bar Q V_{sYM}=0$ implies that 
\be
\g_{mnpqr}^{\a\b} D_\a A_{\b I} = 0,
\label{syml}
\ee
\be
A_{nI} = -{i\over 8}D_\a \g_m^{\a\b} A_{\b I},\quad
W^\b_I = -{i\over{10}}\g^{n\a\b} (  D_\a A_{nI} -\pa_n A_{\a I}),\quad
U_{npI} =
{1\over 8}D_\a (\g_{np})^\a{}_\b W^\b_I = \pa_{[n} A_{p]I}.      
\label{symc}
\ee
Equations (\ref{sugl}) and (\ref{syml}) are
the linearized $N=1$ supergravity and super-Yang-Mills
equations of motion written in terms of the superfields
$A_{\a m}$ and $A_{\a I}$, and equations (\ref{sugc}) and 
(\ref{symc}) define the
linearized supergravity and super-Yang-Mills connections
and field-strengths in terms of $A_{\a m}$ and $A_{\a I}$.
For example, 
the on-shell graviton $h_{nm}$ and gluon $a_{n I}$ are contained in the 
$i(\g^n\th)_\a h_{nm}(x)$ and 
$i(\g^n\th)_\a a_{nI}(x)$ components of $A_{\a m}(x,\th)$ and 
$A_{\a I}(x,\th)$.
The linearized equations of (\ref{sugl})-(\ref{symc})
will be generalized to covariant
non-linear equations in the following subsections.

\subsection{Heterotic superstring in a curved background}

The heterotic sigma model action in a curved background can be
constructed by adding the massless vertex operators of (\ref{vsg})
and (\ref{vsym}) to the flat action of (\ref{flataction}), and then
covariantizing with respect to $N=1$ $D=10$
super-reparameterization invariance. Alternatively, one can
consider the most general action constructed from the worldsheet
variables which is classically invariant under worldsheet conformal
transformations. In addition, for quantum worldsheet conformal
invariance, one needs to include a Fradkin-Tseytlin term which
couples the spacetime dilaton to the worldsheet curvature.

Using the worldsheet variables defined in the previous subsection,
we can write the heterotic sigma model action in the form
\be
S= {1\over {2\pi\a'}} \int d^2 z [\half (G_{MN}(Z)+B_{MN}(Z)) \p Z^M
\bar\p Z^N + E_M^\a(Z) d_\a \bar\p Z^M + \Omega_{M\a}{}^{\b}(Z)
\l^\a w_\b \bar\p Z^M \label{hetsm} 
\ee 
$$ + A_{M I}(Z) \p Z^M \bar
J^I + W_I^\a(Z) d_\a \bar J^I + \half U_{I\a}{}^{\b} \l^\a w_\b \bar
J^I ~+\half \a' \Phi(Z) r +\bar b\p\bar c] + S_\l + S_J $$ 
where
$M=(m,\mu)$ are curved superspace indices, $Z^M=(x^m,\t^\mu)$,
$A=(a,\a)$ are tangent superspace indices, $S_\l$ and $S_J$ are the
same as in the flat action of (\ref{flataction}), $r$ is the
worldsheet curvature, and $[G_{MN},B_{MN},$ $ E_M{}^\a,
\Omega_{M\a}{}^\b,$ $ A_{M I}, W_I^\a,$ $ U_{I\a}{}^{\b}, \Phi]$ are
the background superfields. The ``metric'' $G_{MN}$ is defined in terms of the 
vectorial part of the supervielbein by $G_{MN}=E_N{}^b E_M{}^a\h_{ab}$, and
we shall define $E_A{}^M$ to be the inverse of $E_M{}^A$. 

Ignoring the Fradkin-Tseytlin term $\int d^2 z \Phi(Z)r$, 
(\ref{hetsm}) is the most general action with classical worldsheet
conformal invariance and zero ghost number which can be constructed
from the heterotic worldsheet variables. Note that $d_\a$ carries
conformal weight $(1,0)$, $\l^\a$ carries ghost number +1 and
conformal weight $(0,0)$, and $w_\a$ carries ghost number $-1$ and
conformal weight $(1,0)$.
Since 
the conjugate momentum ghost
variable $w_\a$ can only appear in combinations which preserve the
gauge invariance $\d w_\a= \L^a (\g_a\l)_\a,$
the background
superfields $\Omega_{M\a}{}^\b$ and $U_{I\a}{}^{\b}$ must satisfy
$(\g^{bcde})_\b{}^\a \Omega_{M\a}{}^\b = (\g^{bcde})_\b{}^\a U_{I\a}{}^\b =0$, 
i.e.
\be
\Omega_{M\a}{}^\b= \Omega_M^{(s)} \d_\a^\b +\half \Omega_M{}^{cd}
(\g_{cd})_\a{}^\b, \quad U_{I\a}{}^{\b}= U^{(s)}_I \d_\a^\b +\half
U_I^{cd} (\g_{cd})_\a{}^\b. \label{Omegaalpha} \ee  

It is worthwhile to pause here to say a few words about the geometry of the 
target space which is implied by this action. Clearly, we identify $E_M{}^A$ as 
the usual 
supervielbein matrix, $B_{MN}$ as the two-form potential and $\Phi$ as 
the dilaton. The superfield $A_{MI}$ is the super-Yang-Mills potential 
while the superfields 
$W_I^\a$ and $U_{I\a}{}^\b$ will turn out to be related to the 
spinor and vector super-Yang-Mills field strengths.
The way in which the 
supervielbein enters into the action indicates that the tangent space should be 
a direct sum of bosonic and fermionic subspaces. This is 
different from the structure of the 
tangent space in the Green-Schwarz formalism 
since 
the $E_M{}^\a$
components of the super-vielbein do
not appear in the GS action. So
one only
needs to specify the fermionic subspace of the GS tangent 
space (or, 
dually, the bosonic subspace of the GS cotangent space).
The form of the 
``metric'' $G_{MN}= E_N{}^b E_M{}^a \eta_{ab}$ 
shows that the structure group in the bosonic sector is the Lorentz 
group while the existence of pure spinors implies that the fermionic structure 
group is the 
spin group times scale transformations. At this stage, the 
two Lorentz groups (in the spinor and vector sectors) are independent, although 
later on we shall choose a gauge with respect to one of them after 
which they will become identified. Note also that the spin connection 
$\O_{M\a}{}^\b$ 
appears explicitly in the action. This implies that conventional 
constraints corresponding to tensorial shifts of the connection are restricted 
by the demand that the BRST operator and action be unchanged.

Taking all this into account we find that, in addition to being invariant
under target-space
super-reparameterizations,
the action of (\ref{hetsm}) is invariant under the local
gauge transformations
\be
\d E_M^b = \eta_{cd}\Lambda^{bc} E_M^d, \quad \d E_M{}^{\a}
=\Sigma^\a_\b E^\b_M,\quad \d \Omega_{M\a}{}^\b = \p_M\Sigma_\a^\b
+ \Sigma^\g_\a \Omega_{M\g}{}^\b - \Sigma^\b_\g \Omega_{M\a}{}^\g,
\label{localone} \ee $$\d W^\a_I = \Sigma^\g_\a W^\g_I,\quad \d
U_{I\a}{}^{\b}= \Sigma^\g_\a U_{I \g }{}^\b - \Sigma^\b_\g U_{I\a }{}^\g,
\quad \d \l^\a = \Sigma^\a_\g \l^\g,\quad \d w_\a =
-\Sigma^\g_\a w_\g,$$ where $\Sigma_\a^\b = \Sigma^{(s)}\d_\a^\b
+\half \Sigma^{bc}(\g_{bc})_\a{}^\b$, $\Lambda^{bc}$ and
$\Sigma^{bc}$ parameterize independent local Lorentz
transformations on the vector and spinor indices, and
$\Sigma^{(s)}$ parameterizes local scale transformations on the
spinor indices. Furthermore, the action of (\ref{hetsm}) and the
BRST operator $\l^\a d_\a$ are invariant under the local shift
transformations
\be
\d \Omega_\a^{(s)}  = (\g_c)_{\a\b} h^{c\b},\quad \d\Omega_\a^{bc}=
2(\g^{[b})_{\a\b} h^{c]\b},\quad \d d_\a = -\d\O_{\a\b}{}^\g \l^\b
w_\g,\quad \d U_{I\a}{}^{\b}=W_I^\g \d \O_{\g\a}{}^\b, \label{wgauge}
\ee where $\Omega_{\a\b}{}^\g=E_\a^M \Omega_{M\b}{}^\g$, $h^{c\d}$
is a local gauge parameter, and the transformation of
$\O_{\a\b}{}^\g$ has been chosen such that $\l^\a \d d_\a=0$. Note
that $d_\a$ can be treated as an independent variable in the action
of (\ref{hetsm}) since $p_\a$ does not appear explicitly.

The first term in the first and second line of (\ref{hetsm}) is the
standard heterotic GS action, but the other terms will be needed for
BRST invariance, just as in the linearized vertex operators
of (\ref{vsg}) and (\ref{vsym}). 
As will now be shown to lowest order in $\a'$,
nilpotence and holomorphicity of $\l^\a d_\a$ implies the equations
of motion for the background superfields in (\ref{hetsm}). Note
that nilpotence and antiholomorphicity of the right-moving BRST
current, $\bar c\bar T + \bar c\bar\p\bar c\bar b$, does not impose
any conditions to lowest order in $\a'$ because the action of (\ref{hetsm})
is classically conformally invariant.

\subsection{Heterotic nilpotency constraints}

We shall first derive the constraints coming from nilpotency of
$Q=\oint\l^\a d_\a$. Defining the canonical momentum $P_M$ in the usual
manner as $P_M = \p L/\p(\p_0 Z^M)$, one finds that
\be
d_\a = E_\a^M [P_M +\half B_{MN}(\p Z^N - \bar\p Z^N) -
\O_{M\a}{}^{\b} \l^\a w_\b - A_{M I} \bar J^I ]. \label{canm} \ee
Using the canonical commutation relations 
\be
[P_M, Z^N ] =\d_M^N,\quad [w_\a,\l^\b] = \d_\a^\b,\quad [\bar
J^I,\bar J^J] = f^{IJ}_K \bar J^K, \label{canonical} \ee one
computes that
\be
\{Q,Q\} = \oint\l^\a \l^\b [T_{\a\b}{}^C D_C + \half H_{\a\b M}(\p
Z^M - \bar\p Z^M) - R_{\a\b\g}{}^{\d} \l^\g w_\d - F_{\a\b I}\bar
J^I] \label{putt} \ee where $D_C = E_C^M (P_M - \O_{M\a}{}^\b \l^\a
w_\b - A_{M I} \bar J^I)$. The torsions $T_{AB}{}^C$, three-form
$H_{ABC}$, curvatures $R_{AB\g}{}^\d$, and field strengths $F_{AB
I}$ in (\ref{putt}) are defined by
\be
[\nab_A,\nab_B] = T_{AB}{}^C \nab_C + R_{AB}^{(s)} S +
R_{AB}{}^{ab} M_{ab} + F_{AB~I} Y^I, \quad H_{ABC} =3 E_A^M E_B^N
E_C^P \p_{[M} B_{NP]}, \label{conn} \ee where $\nab_A = E_A^M (\p_M
+ \O_M^{(s)} S+ \O_M^{ab} M_{ab} + A_{M I} Y^I)$, $S$ is a
scale generator which transforms $\d E_\a{}^M=\Lambda^{(s)} E_\a{}^M$,
$M_{ab}$ is the Lorentz generator, $Y^I$ is the gauge group
generator and $R_{AB\b}{}^\g= R_{AB}^{(s)}+\half
R_{AB}{}^{cd}(\g_{cd})_\b{}^\g$. Note that  $f_{[AB]}$
signifies the graded commutator, i.e. $f_{[AB]}=\half(f_{AB}+f_{BA})$ 
when both indices are
fermionic and $f_{[AB]}=\half(f_{AB}-f_{BA})$ otherwise.

So nilpotency of $Q$ implies the constraints
\be
\l^\a \l^\b T_{\a\b}{}^C =\l^\a \l^\b  H_{\a\b B} =\l^\a \l^\b
\l^\g R_{\a\b\g}{}^\d = \l^\a\l^\b F_{\a\b I} =0 \label{nilphet}
\ee for any $\l^\a$ satisfying the pure spinor constraint of
(\ref{pure}).
Note that the $\l^\a\l^\b\l^\g R_{\a\b\g}{}^\d=0$ constraint
is implied by $\l^\a\l^\b T_{\a\b}{}^C=0$ through Bianchi identities.

As shown in \cite{h}, the constraints (\ref{nilphet})
follow from pure spinor integrability in loop superspace and imply
all the essential 
N=1 supergravity/super-Yang-Mills constraints. 
Indeed, the chirality operator introduced in \cite{h} 
in pure spinor loop superspace precisely coincides with the BRST operator $Q$.
So (\ref{nilphet}) implies all but the ``conventional''
constraints which define the vector components of superfields in
terms of their spinor components and define the spin connection in
terms of the super-vierbein. As will be shown below, these
conventional constraints (up to gauge invariances) are implied
by the holomorphicity of $\l^\a d_\a$. 
\subsection{ Heterotic holomorphicity constraints }

We shall now derive the constraints coming from holomorphicity of
$\l^\a d_\a$. Varying $\l^\a$ and its conjugate momentum in
(\ref{hetsm}) and ignoring the contribution from the
Fradkin-Tseytlin term which is higher order in $\a'$, one obtains the equations
\be
\bar\p\l^\a = -(\Omega_{M\b}{}^\a \bar\p Z^M + U_{I\b }{}^\a \bar
J^I) \l^\b, \quad \bar\p w_\a = (\Omega_{M\a}{}^\b \bar\p Z^M +
U_{I\a}{}^{\b} \bar J^I)  w_\b, \label{dbarl} \ee and varying the
right-moving variables, one obtains the equations
\be
\p \bar J^K = f_J^{IK} (A_{M I} \p Z^M  + W_I^\a d_\a + U_{I \a }{}^{\b} 
\l^\a w_\b) \bar J^J, \ee where $f_J^{IK}$ are the Lie
algebra structure constants. And by varying $d_\a$, one obtains the
equation of motion
\be
E_M{}^{\a} \bar\p Z^M =  - W^\a_I \bar J^I. \label{dbarZ} \ee 
Finally,
by computing $E_\a^P (\d S/\d Z^P)$, one obtains the equation of
motion
\be
\bar \p d_\a = E_\a^P [  (\p_{[P} E_{M]}^a E_N^b \eta_{ab}+
\p_{[P} E_{N]}^a E_M^b \eta_{ab} + \half
H_{PMN}) \p Z^M \bar\p Z^N \label{barpd}
\ee
$$ + 2(\p_{[P} E_{N]}^\b d_\b + \p_{[P} \O_{N]\g}{}^\b \l^\g
w_\b)\bar\p Z^N - \O_{P\g}{}^\b\bar\p (\l^\g w_\b)
 - A_{P I} \p\bar J^I
$$ $$+ (2\p_{[P} A_{M] I}\p Z^M +\p_P W_I^\b d_\b +\p_P U_{I\g}{}^{\b}
\l^\g w_\b) \bar J^I]. $$

Putting these equations together, one finds
\be
\bar\p(\l^\a d_\a) = \l^\a[\half( T_{\a bc} + T_{\a cb} + H_{\a
bc}) \Pi^b + \half (T_{\a \b c} + H_{\a \b c}) \Pi^\b + T_{\a
c}{}^\b d_\b  +  R_{\a c \b}{}^{\g} \l^\b w_\g]\bar\Pi^c
\label{putwo} \ee $$+ \l^\a [(F_{\a b I} -\half W_I^\b ( T_{\a\b b}
+ H_{\a b\b})) \Pi^b + (F_{\a\g I}-\half W_I^\b H_{\a\g\b})
\Pi^\g]\bar J^I$$ $$ + \l^\a [(\nab_\a W_I^\b - T_{\a\g}{}^\b
W_I^\g - U_{I\a}{}^{\b} )d_\b + (\nab_\a U_{\g I}^\d  - 
R_{\a\b\g}{}^\d W_I^\b)\l^\g w_\d] \bar J^I,$$ where $\Pi^A = E^A_M
\p Z^M$, $\bar\Pi^A = E^A_M \bar\p Z^M$ and $T_{ABc}= T_{AB}{}^d
\eta_{cd}$.

So from (\ref{putwo}), $\bar\p(\l^\a d_\a)=0$ implies the
constraints
\be
T_{\a(ab)} = H_{\a ab} =  T_{\a\b c} + H_{\a\b c} = T_{\a c}{}^\b =
 0, \quad \l^\a\l^\b R_{\a c \b}{}^\g=0,
\quad F_{\a\b I}= \half W_I^\g H_{\a\b\g}, \label{holohet} \ee
$$F_{\a b I} = W_I^\b T_{\a\b b},\quad \nab_\a W^\b_I
-T_{\a\g}{}^\b W_I^\g =  U_{I\a}{}^{\b},\quad \l^\a\l^\b
(\nab_\a U_{I\b }{}^\d - R_{\a\g\b}{}^\d W_I^\g)=0,$$ where
$\l^\a$ is any spinor satisfying (\ref{pure}).

The constraints of (\ref{nilphet}) and (\ref{holohet}) will now be
shown to imply the correct supergravity and super-Yang-Mills
equations of motion.

\subsection{N=1 supergravity/super-Yang-Mills constraints}

It will be useful to first consider the supergravity constraints of
(\ref{nilphet}) and (\ref{holohet}) which have lowest scaling
dimension since the higher dimensional constraints will be implied
by these constraints through Bianchi identities. At dimension
$-\half$, the only constraint is $\l^\a\l^\b H_{\a\b\g}=0$ which
implies that $H_{\a\b\g}=0$ since there is no non-zero symmetric
$H_{\a\b\g}$ satisfying $\l^\a\l^\b H_{\a\b\g}=0$. 

At dimension 0,
the constraints $\l^\a\l^\b T_{\a\b}{}^c= \l^\a\l^\b H_{\a\b c}=0$
and $T_{\a\b}{}^c= -\eta^{cd} H_{\a\b d}$ imply that $T_{\a\b}{}^c =
-\eta^{cd} 
H_{\a\b d}= i(\g^d)_{\a\b} f_d^c$ for some $f_d^c$. The dimension zero 
Bianchi identity $D_{(\a}H_{\b\g\d)}=
T_{(\a \b}^D H_{\g\d )D}$ then tells us that $f_d^c$ is an SO(9,1)
matrix times a scale factor. So 
using the local spinor Lorentz and scale transformations of
(\ref{localone}), $f_d^c$ can be gauge-fixed to $\d_d^c$ so that
\be
T_{\a\b}{}^c = -\eta^{cd} H_{\a\b d} =i(\g^c)_{\a\b}. \label{scalars}
\ee
Note that at this point we still have one local Lorentz symmetry, acting now on 
both spinor and vector indices. The connection for this symmetry is 
$\O_{M}{}^{ab}$. On the other hand, the fermionic scale invariance has been 
fixed and so it 
need not be the case that other components of the torsion should 
respect this symmetry.

At dimension $\half$, the constraint $\l^\a\l^\b T_{\a\b}{}^\g=0$
implies that $T_{\a\b}{}^\g = f_c^\g (\g^c)_{\a\b}$ for some
$f_c^\g$. Using the shift symmetry of (\ref{wgauge}), $f_c^\g$ can
be gauge-fixed to zero so that $T_{\a\b}{}^\g=0$. The other
dimension $\half$ constraints, $H_{\a c d}=T_{\a (cd)}=0$, imply
through the Bianchi identity $\nabla_{(\a} T_{\b\g)}{}^c = -T_{(\a
\b}{}^D T_{\g) D}{}^c$ that $T_{\g b}{}^c=
2\eta^{cd}(\g_{bd})_\a{}^\b \O^{(s)}_\b$.

At dimension one, the constraint $T_{c \a}{}^\b=0$ decomposes into
\be
T_{c\a}{}^\b= T_c^{defg}(\g_{defg})_\a{}^\b+
T_c^{de}(\g_{de})_\a{}^\b + T_c \d_\a^\b=0. \label{decomp} \ee The
constraints $T_c=0$ and $T_c^{de}=0$ determine the vector
components of the spin connections $\O_c^{(s)}$ and $\O_c{}^{de}$,
whereas the constraint $T_c^{defg}=0$ is implied by the Bianchi
identity $(\nabla H + TH)_{bc\alpha\gamma} (\g^{bdefg})^{\a\g}=0$.
Similarly, the constraints involving the curvature tensor in
(\ref{nilphet}) and (\ref{holohet}) are implied by the Bianchi
identity $R_{[ABC]}{}^D=\nabla_{[A}T_{BC]}{}^D+T_{[AB}{}^E
T_{C]E}{}^F$.

To extract the supergravity equations unambiguously from the above constraints 
it is convenient to reduce the structure group from Lorentz group times 
fermionic scale to just the Lorentz group. The dimension zero torsions are 
unchanged but the dimension one-half torsion $T_{\a\b}{}^\c$ gets amended to
\be
T_{\a\b}{}^\c\rightarrow 
T_{\a\b}{}^\g - 2 \d_{(\a}{}^\g \O_{\b)}^{(s)} = -
2\d_{(\a}{}^\c \O^{(s)}_{\b)}.
\ee

There are corresponding changes at higher dimensions. The leading component of 
$\O^{(s)}_\a$ is the dilatino and to show that there are no unwanted fields
one 
must show that this superfield is proportional to
the spinorial derivative of a scalar 
superfield $\Phi$ whose leading 
component is the dilaton. It is straightforward to 
verify that this is the case, although it is necessary to go to dimension 
three-halves to do so.
As discussed in section 5, holomorphicity of $\l^\a d_\a$ to the next
order in $\a'$ will imply that this scalar superfield $\Phi$ is the
same superfield that appears in the Fradkin-Tseytlin term of (\ref{hetsm}).

The above supergravity constraints therefore imply that all of the supergravity
superfields can be expressed in terms of the spinor supervielbein
$E_\a{}^M$, and the equation $T_{\a\b}{}^c=i(\g^c)_{\a\b}$ puts $E_\a^M$
on-shell. Similarly, the super-Yang-Mills constraints in
(\ref{nilphet}) and (\ref{holohet}) imply that the super-Yang-Mills
superfields $A_{c I}$, $W_{I}^\a$ and $U_{I\a}{}^{\b}$ can be
expressed in terms of the spinor superfield $A_{\a I}$, and the
equation $F_{\a\b I}=0$ puts $A_{\a I}$ on-shell. So nilpotence and
holomorphicity of $\l^\a d_\a$ has been shown to imply the N=1
supergravity/super-Yang-Mills equations of motion to lowest order
in $\a'$.

\section{Type II Superstring Sigma Model}

In this section, the pure spinor version of the Type IIA and IIB superstring
will be reviewed in flat and curved backgrounds. Nilpotence and
holomorphicity of $\l^\a d_\a$ and nilpotence and antiholomorphicity
of $\hl^\ah d_\ah$ will then be shown to imply the
superspace equations of motion for the 
N=2 supergravity background.

\subsection{Type II superstring in a flat background}

In the pure spinor version of the Type II superstring, the worldsheet
variables consist of the $N=2$ $D=10$
superspace variables $(x^m,\th^\a,p_\a,\hat\t^\ah,\hat p_\ah)$ for 
$m=0$ to 9 and
$\a,\ah=1$ to 16 where $p_\a$ is the conjugate momentum to $\th^\a$
and $\hat p_\ah$ is the conjugate momentum to $\hht^\ah$. For
the Type IIA superstring, $\a$ and $\ah$ denote SO(9,1) spinors of opposite
chirality while for the Type IIB superstring, $\a$ and $\ah$ denote
SO(9,1) spinors of the same chirality. The pure spinor formalism
also contains the worldsheet variables $\l^\a$ and $\hl^\ah$, and
their conjugate momenta $w_\a$ and $\hw_\ah$,
which are constrained to satisfy the pure spinor conditions 
\be
\l\g^m\l=0, \quad \hl\g^m\hl=0
\label{pureII}
\ee
for $m=0$ to 9.
In a flat
background, $\t^\a$, $p_\a$, $\l^\a$ and $w_\a$ 
are left-moving while $\hht^\ah$,
$\hp_\ah$, $\hl_\ah$ and $\hw_\ah$ are right-moving.

The action and stress-tensor in a flat background is
\be
S={1\over{2\pi \a'}}
\int d^2 z (\half \pa x^m \bar\pa x_m + p_\a \bar\pa\th^\a 
+\hp_\ah \p\hht^\ah) +S_\l + S_\hl ,
\label{flatactionII}
\ee
\be
T ={1\over{\a'}}( -\half \pa x^m \pa x_m -p_\a \pa\t^\a) + T_\l,\quad
\bar T = {1\over{\a'}}(-\half \bar\pa x^m \bar\pa x_m -\hp_\ah\bar \pa\hht^\ah)
 + \bar T_\hl,
\ee
where $S_\l$ and $S_\hl$ are the actions for $\l^\a$ and $\hl^\ah$,
and $T_\l$ and $\bar T_\hl$ 
are the $c=22$ left and right-moving stress tensors for
$\l^\a$ and $\hl^\ah$.
As in the heterotic case, the explicit form of $S_\l$ and $S_\hl$ will
not be needed. We will only need to know that one can construct
left and right-moving Lorentz currents, $N^{mn}={1\over{2\a'}} \l\g^{mn} w$
and $\hat N^{mn}={1\over{2\a'}} \hl\g^{mn}\hw$, which satisfy the OPE's
\be
N^{mn}(y) \l^\a (z) \to
\half(\g^{mn})^\a{}_\b {{\l^\b(z)}\over{y-z}}, \quad
\hat N^{mn}(\bar y) \hl^\ah (\bar z) \to
\half(\g^{mn})^\ah{}_\bh {{\hl^\bh(\bar  z)}\over{\bar y-\bar z}},
\label{OPElII}
\ee
\be
N^{kl}(y) N^{mn}(z) \to
{{\eta^{m[l} N^{k]n}(z) -
\eta^{n[l} N^{k]m}(z) }\over {y-z}} - 3
{{\eta^{kn} \eta^{lm} -
\eta^{km} \eta^{ln}}\over{(y-z)^2}}  .
\label{OPENII}
\ee
\be
\hat N^{kl}(\bar y)\hat N^{mn}(\bar z) \to
{{\eta^{m[l} \hat N^{k]n}(\bar z) -
\eta^{n[l} \hat N^{k]m}(\bar z) }\over {\bar y-\bar z}} - 3
{{\eta^{kn} \eta^{lm} -
\eta^{km} \eta^{ln}}\over{(\bar y-\bar z)^2}}  .
\ee

Physical states of the superstring are defined as vertex operators
in the cohomology of the left and right-moving BRST operators
\be
Q= \oint \l^\a d_\a,\quad
\bar Q = \oint \hl^\ah \hd_\ah
\ee
where 
\be
d_\a = p_\a -{i\over 2}\g_{\a\b}^m \t^\b \p x_m +{1\over 8}
\g_{\a\b}^m \g_{m~\g\d}\t^\b\t^\g\p\t^\d, \quad 
\Pi^m = \p x^m +{i\over 2}\t\g^m\p\t
\ee
\be
\hd_\ah = \hp_\ah -{i\over 2}\g_{\ah\bh}^m \hht^\b \bar\p x_m +{1\over 8}
\g_{\ah\bh}^m \g_{m~\gh\dh}\hht^\bh\hht^\gh\bar\p\hht^\dh, \quad 
\bar\Pi^m = \bar\p x^m +{i\over 2}\hht\g^m\bar\p\hht
\ee
are spacetime supersymmetric and satisfy the OPE's
\be
d_\a(y) d_\b(z) \to -i\a'(y-z)^{-1} \g_{\a\b}^m \Pi_m(z),\quad
d_\a(y) \Pi^m(z) \to  i\a'(y-z)^{-1} \g_{\a\b}^m \p\t^\b(z),
\label{OPEdII}
\ee
\be
\hd_\ah(\bar y) \hd_\bh(\bar z) \to -i
\a'(\bar y-\bar z)^{-1} \g_{\ah\bh}^m \bar 
\Pi_m(\bar z),\quad
\hd_\ah(\bar y) \bar\Pi^m(\bar z) \to i\a' 
(\bar y-\bar z)^{-1} \g_{\ah\bh}^m \bar\p\hht^\bh(\bar z),
\ee

To construct the sigma model for the Type II superstring,
it will be useful to know the integrated form of the massless
Type II supergravity vertex operator
\be
V_{SG} = \int d^2 z [
\pa\th^\a \bar\p\hht^\bh A_{\a \bh}(x,\t,\hht) + 
\pa\th^\a \bar\Pi^m A_{\a m}(x,\t,\hht) + 
\Pi^m \bar\p\hht^\ah A_{m \ah}(x,\t,\hht)
+\Pi^m \bar\Pi^n A_{m n}(x,\t,\hht)
\label{vsgII}
\ee
$$+ d_\a (\bar\p\hht^\bh E^\a_\bh(x,\t,\hht) + \bar\Pi^m E^\a_m(x,\t,\hht)) 
+ \hd_\ah (\p\t^\b E^\ah_\b (x,\t,\hht)+ \Pi^m E^\ah_m(x,\t,\hht)) $$
$$ +\half N_{mn}
 (\bar\p\hht^\bh \O^{mn}_\bh(x,\t,\hht) + \bar\Pi^p\O^{mn}_p(x,\t,\hht)) 
+\half \hat N_{mn}
(\p\t^\b \hat\O^{mn}_\b (x,\t,\hht)+ \Pi^p \hat\O^{mn}_p(x,\t,\hht)) $$
$$+ d_\a \hd_\bh P^{\a\bh}(x,\t,\hht) + N_{mn} \hd_\ah C^{mn\ah}(x,\t,\hht)
+ d_\a \hat N_{mn} \hat C^{\a mn} (x,\t,\hht)
+ N_{mn} \hat N_{pq} S^{mnpq}(x,\t,\hht)].$$
Note that the first line of $V_{SG}$ is the
same as in the Green-Schwarz Type II superstring vertex operator,
but the other lines are needed for the vertex operator
to be BRST invariant. The Type II superstring 
vertex operator of (\ref{vsgII}) can
be understood as the ``square'' of the open superstring vertex
operator of (\ref{open}).

Using (\ref{pureII}) and
the OPE's of (\ref{OPElII}) and (\ref{OPEdII}), one can check
that $QV_{SG} = \bar Q V_{SG}=0$ implies that 
\be
\g_{mnpqr}^{\a\b} D_\a A_{\b \gh} = 0,\quad
\g_{mnpqr}^{\ah\bh} \hat D_\ah A_{\g \bh} = 0,
\label{suglII}
\ee
\be
A_{n\gh} = -{i\over 8}D_\a \g_n^{\a\b} A_{\b \gh},\quad
A_{\g n} = -{i\over 8}\hat D_\ah \g_n^{\ah\bh} A_{\g \bh},\quad
A_{m n} = {1\over {64}}D_\a\hat D_\gh \g_m^{\a\b}\g_n^{\gh\dh} A_{\b \dh},
\label{sugcII}
\ee
and similar equations for the superfields $(E,\O,
\hat\O,P,C,\hat C,S)$ in terms of $A_{\a\bh}$.
Note that $D_\a = {\p\over{\p\t^\a}} +{i\over 2} \g^m_{\a\b}\t^\b\p_m$ and
$\hat D_\ah = {\p\over{\p\hht^\ah}} +{i\over 2} \g^m_{\ah\bh}\hht^\bh\p_m$ are
the N=2 D=10 supersymmetric derivatives.
Equations (\ref{suglII}) are
the linearized $N=2$ supergravity 
equations of motion written in terms of the superfield
$A_{\a\bh}$, and equations (\ref{sugcII}) 
define the
linearized supergravity
connections
in terms of $A_{\a \bh}$.
For example, 
the on-shell graviton $h_{nm}$ is contained in the 
$(\g^n\th)_\a (\g^m\hht)_\bh h_{nm}(x)$ of
$A_{\a \bh}(x,\th,\hht)$.
These linearized equations will be generalized to covariant
non-linear equations in the following subsections.

\subsection{Type II superstring in a curved background}

The Type II sigma model
action in a curved background (except for the Fradkin-Tseytlin
term) can be constructed by
adding the massless vertex operator of 
(\ref{vsgII})
to the flat action of (\ref{flatactionII}), and then covariantizing
with respect to $N=2$ $D=10$ super-reparameterization invariance.
Alternatively, one can consider the most general action constructed
from the worldsheet variables which 
is classically invariant under
worldsheet conformal transformations. 

Using 
the worldsheet variables defined in the previous subsection, we can write the
Type II sigma model action as
\be
S= {1\over {2\pi\a'}} \int d^2 z [\half (G_{MN}(Z)+B_{MN}(Z)) \p Z^M \bar\p Z^N
\label{IIsm}
\ee
$$
+ E_M^\a(Z) d_\a \bar\p Z^M + 
E_M^\ah(Z) \hd_\ah \p Z^M 
+ \Omega_{M\a}{}^\b(Z) \l^\a w_\b \bar\p Z^M
+ \hat\Omega_{M\ah}{}^\bh(Z)\hl^\ah \hat w_\bh \p Z^M
+ P^{\a\bh}(Z) d_\a \hd_\bh
$$
$$+ C_{\a}^{\b\gh}(Z) \l^\a w_\b \hd_\gh 
+\hat C_{\ah}^{\bh\g}(Z) \hl^\ah \hat w_\bh  d_\g +
S_{\a\gh}^{\b\dh}(Z) \l^\a w_\b \hl^\gh \hat w_\dh ~+\half\a' \Phi(Z) r ] 
+ S_\l + S_\hl$$
where $M=(m,\mu,\hat\mu)$ are curved superspace indices, $Z^M=(x^m,\t^\mu,
\hht^{\hat\mu})$,
$A=(a,\a,\ah)$
are tangent superspace indices, $S_\l$ and $S_\hl$ are the same as
in the flat action of (\ref{flatactionII}), $r$ is the worldsheet
curvature, and
$[G_{MN}=\eta_{cd} E_M^c E_N^d,B_{MN},$
$ E_M^\a, E_M^\ah,\O_{M\a}{}^\b,\hat \Omega_{M\ah}{}^\bh,$
$P^{\a\bh}, C_{\a}^{\b\gh},\hat C_{\ah}^{\bh\g}, S_{\a\gh}^{\b\dh},\Phi]$
are the background superfields. 

If the Fradkin-Tseytlin term, $\int d^2 z \Phi(Z)r$, is omitted
(\ref{IIsm}) is the most general action with classical worldsheet
conformal invariance and zero (left,right)-moving ghost number 
which can be constructed
from the Type II worldsheet variables. Note that $d_\a$ carries
conformal weight $(1,0)$, $\hat d_\ah$ carries conformal weight
$(0,1)$, $\l^\a$ carries ghost number $(1,0)$ and
conformal weight $(0,0)$, 
$\hl^\ah$ carries ghost number $(0,1)$ and
conformal weight $(0,0)$, 
$w_\a$ carries ghost number $(-1,0)$ and
conformal weight $(1,0)$, and
$\hat w_\ah$ carries ghost number $(0,-1)$ and
conformal weight $(0,1)$.
Since 
$w_\a$ and $\hat w_\ah$ can only appear in combinations which commute with
the pure spinor constraints of (\ref{pureII}), the background superfields
must satisfy 
\be
(\g^{bcde})_\b^\a
\Omega_{M\a}{}^\b = 
(\g^{bcde})_\b^\a
\hat\Omega_{M\ah}{}^\bh = 
(\g^{bcde})_\b^\a C_{\a}^{\b\gh}
=(\g^{bcde})_\bh^\ah \hat C_{\ah}^{\bh\g} =
(\g^{bcde})_\b^\a S_{\a\gh}^{\b\dh} =
(\g^{bcde})_\dh^\gh S_{\a\gh}^{\b\dh} = 0,
\label{conbaII}
\ee
and the different components of the spin connections will be defined as
\be
\Omega_{M\a}{}^\b= \Omega_M^{(s)} \d_\a^\b +\half
\Omega_M^{cd} (\g_{cd})_\a{}^\b,
\quad
\hat\Omega_{M\ah}{}^\bh=\hat\Omega_M^{(s)} \d_\ah^\bh +\half
\hat\Omega_M^{cd} (\g_{cd})_\ah{}^\bh. 
\label{OmegaII}
\ee

Although the background superfields appearing in (\ref{IIsm}) look
unconventional, they all have physical interpretations. The superfields
$E_M{}^A$, $B_{MN}$ and $\Phi$ are the supervielbein, two-form potential
and dilaton superfields, $P^{\a\bh}$ is the superfield whose lowest
components are the Type II Ramond-Ramond field strengths, and
the superfields
$C_\a^{\b\gh}=C^\gh \d_\a^\b +\half C^{\gh ab}(\g_{ab})_\a^\b$ and
$\hat C_\ah^{\bh\g}=\hat C^\g \d_\ah^\bh +\half
 \hat C^{\g ab}(\g_{ab})_\ah^\bh$
are related to the N=2 D=10 dilatino and gravitino field strengths.
As in the heterotic sigma model, the form of the metric in the
Type II sigma model implies
that the structure group in the bosonic sector is the Lorentz group.
But there are now two independent pure spinors, so one has two
independent fermionic structure groups, each consisting of the 
spin group times scale transformations. One therefore has two
independent sets of spin connections and scale connections,
$(\Omega_M^{(s)},\Omega_M^{ab})$
and $(\hat\Omega_M^{(s)},\hat\Omega_M^{ab})$, which appear explicitly
in the Type II sigma model action. Finally, the background superfields
$S_{\a\gh}^{\b\dh}$ appearing in (\ref{IIsm}) will be related to
curvatures constructed from these spin and scale connections.
Note that a similar relation occurs in the RNS sigma model action
which contains the terms 
\be
{1\over{4\pi\a'}}\int d^2 z(\Omega_m^{ab}(x)\psi_a\psi_b \bar\p x^m
+\hat\Omega_m^{ab}(x)\bar\psi_a\bar\psi_b \p x^m + 
S_{abcd}(x) \psi^a\psi^b\bar\psi^c\bar\psi^d)
\label{RNSsm}
\ee
where $\psi^a= e^a_m(x)\psi^m$, $\bar\psi^a= e^a_m(x)\bar\psi^m$,
and $e_m^a(x)$ is the target-space vielbein.

In addition to being target-space super-reparameterization invariant,
the action of (\ref{IIsm}) is invariant under the local
gauge transformations
\be
\d E_M^b =
\eta_{cd}\Lambda^{bc} E_M^d, \quad \d E^\a_M =\Sigma^\a_\b E^\b_M,\quad
\d E^\ah_M =\hat\Sigma^\ah_\bh E^\bh_M,
\label{localtwo}
\ee
$$
\d \Omega_{M\a}{}^\b = \p_M\Sigma_\a^\b +
\Sigma^\g_\a \Omega_{M\g}{}^\b -
\Sigma^\b_\g \Omega_{M\a}{}^\g,
\quad \d \hat\Omega_{M\ah}{}^\bh = \p_M\hat\Sigma_\ah^\bh +
\hat\Sigma^\gh_\ah \hat\Omega_{M\gh}{}^\bh -
\hat\Sigma^\bh_\gh \hat\Omega_{M\ah}{}^\gh,$$
$$\d \l^\a =
\Sigma^\a_\g \l^\g,\quad
\d w_\a =
-\Sigma^\g_\a w_\g, \quad
\d \hl^\ah =
\hat\Sigma^\ah_\gh \hl^\gh,\quad
\d \hat w_\ah =
-\hat\Sigma^\gh_\ah w_\gh,$$
where $\Sigma_\a^\b
= \Sigma^{(s)}\d_\a^\b +\half \Sigma^{bc}(\g_{bc})_\a{}^\b$, 
$\hat\Sigma_\ah^\bh
= \hat\Sigma^{(s)}\d_\ah^\bh +\half \hat\Sigma^{bc}(\g_{bc})_\ah{}^\bh$, 
$[\Lambda^{bc} ,\Sigma^{bc},\hat\Sigma^{bc}]$
parameterize independent local Lorentz transformations
on the [vector, unhatted spinor, hatted spinor] indices, $\Sigma^{(s)}$
and $\hat\Sigma^{(s)}$
parameterize independent local scale transformations on the unhatted and
hatted spinor indices, and the background superfields $[P^{\a\ah},
C_\a^{\b\gh},\hat C_{\ah}^{\bh\g}, S_{\a\gh}^{\b\dh}]$ transform
according to their spinor indices.

Furthermore, the action of (\ref{IIsm}) and the BRST operators
$\l^\a d_\a$ and $\hl^\ah \hd_\ah$
are invariant under the local shift transformations
\be
\d \Omega_\a^{(s)}  = (\g_c)_{\a\b} h^{c\d},\quad
\d\Omega_\a^{bc}= 2 (\g^{[b})_{\b\d} h^{c]\d},\quad
\d d_\a = -\d\O_{\a\b}{}^\g \l^\b w_\g,
\label{wgaugeII}
\ee
$$\d \hat\Omega_\ah^{(s)}  = (\g_c)_{\ah\bh} \hat h^{c\dh},\quad
\d\hat\Omega_\ah^{bc}= 2(\g^{[b})_{\bh\dh} \hat h^{c]\d},\quad
\d \hd_\a = -\d\hO_{\ah\bh}{}^\gh \hl^\bh \hat w_\gh,$$
$$\d C_\a^{\b\gh} = P^{\d\gh} \d\O_{\d\a}{}^\b,\quad
\d \hat C_\ah^{\bh\g} = -P^{\g\dh} \d\hO_{\dh\ah}{}^\bh,\quad
\d S_{\a\gh}^{\b\dh}= \hat C_\gh^{\dh\kappa} \d\O_{\kappa\a}{}^\b
+ C_\a^{\b\hat\kappa} \d\hO_{\hat\kappa\gh}{}^\dh$$
where $h^{c\d}$ and $\hat h^{c\dh}$ are independent local
gauge parameters and the transformations of $\O_{\a\b}{}^\g$
and $\hO_{\ah\bh}{}^\gh$ have
been chosen such that $\l^\a \d d_\a=\hl^\ah \d d_\ah =0$.
Note that $d_\a$ and $\hd_\ah$
can be treated as independent variables in (\ref{IIsm})
since $p_\a$ and $\hat p_\ah$ do not appear explicitly.

The first line
of (\ref{IIsm})
is the standard Type II GS action, but the other lines
are needed for BRST invariance. As will now be shown to lowest order
in $\a'$, nilpotence and holomorphicity
of $\l^\a d_\a$ and nilpotence and antiholomorphicity
of $\hl^\ah \hd_\ah$ imply the 
equations of motion for the background superfields
in (\ref{IIsm}).

\subsection{Type II nilpotency constraints}

To analyze the conditions implied by nilpotency of
$Q=\oint  \l^\a d_\a$ and $\hat Q=\oint
\hl^\ah \hd_\ah$, it is convenient to use
the canonical momenta
$P_M = \p L/ \p (\p_0 Z^M)$ to write
\be
d_\a = E_\a^M [P_M + \half
B_{M N} (\p Z^N -\bar\p Z^N) - \O_{M\b}{}^\g \l^\b w_\g
- \hO_{M\bh}{}^\gh \hl^\bh \hat w_\gh],
\label{defnilII}
\ee
$$\hd_\ah = E_\ah^M [P_M + \half 
B_{M N} (\p Z^N -\bar\p Z^N) - \O_{M\b}{}^\g \l^\b w_\g
- \hO_{M\bh}{}^\gh \hl^\bh \hat w_\gh].$$

Using the canonical commutation relations  
$$[P_M,Z^N\}=\d_M^N,\quad [w_\a, \l^\b] = \d_\a^\b,\quad 
[\hat w_\ah, \hat \l^\bh] = \d_\ah^\bh,$$
one finds that
$$\{Q,Q\}= \oint \l^\a\l^\b [T_{\a\b}{}^C D_C + 
\half (\p Z^N-\bar\p Z^N)H_{\a\b N}
- R_{\a\b\g}{}^\d \l^\g w_\d - \hat R_{\a\b\gh}{}^\dh \hl^\gh \hat w_\dh],$$
$$\{\hat Q,\hat Q\}= \oint \hl^\ah\hl^\bh [T_{\ah\bh}{}^C D_C + 
\half (\p Z^N-\bar\p Z^N)H_{\ah\bh N}
- R_{\ah\bh\g}{}^\d 
\l^\g w_\d - \hat R_{\ah\bh\gh}{}^\dh \hl^\gh \hat w_\dh],$$
$$\{Q,\hat Q\}= \oint \l^\a\hl^\bh [T_{\a\bh}{}^C D_C + 
\half (\p Z^N-\bar\p Z^N)H_{\a\bh N}
- R_{\a\bh\g}{}^\d \l^\g w_\d - \hat R_{\a\bh\gh}{}^\dh \hl^\gh \hat w_\dh],$$
where $D_C = E_C^M (P_M - \O_{M\a}{}^\b \l^\a w_\b 
 - \hat \O_{M\ah}{}^\bh \hl^\ah \hat w_\bh )$, 
$T_{AB}{}^\a$ and $R_{AB\b}{}^\g$ are defined 
using the $\Omega_{M\b}{}^\g$ spin connection,
and $T_{AB}{}^\ah$ and $\hat R_{AB\bh}{}^\gh$ are defined 
using the $\hat\Omega_{M\bh}{}^\gh$ spin connection.

So nilpotency of $Q$ and $\hat Q$ implies that 
\be
\l^\a \l^\b T_{\a\b}{}^C = \l^\a \l^\b H_{\a\b B}= \l^\a\l^\b
\hat R_{\a\b\gh}{}^\dh = \l^\a\l^\b\l^\g R_{\a\b\g}{}^\d =0,
\label{nilII}
\ee
$$\hl^\ah \hl^\bh T_{\ah\bh}{}^C = \hl^\ah \hl^\bh H_{\ah\bh B}= \hl^\ah\hl^\bh
\hat R_{\ah\bh\g}{}^\d = \hl^\a\hl^\b\hl^\gh R_{\ah\bh\gh}{}^\dh =0,$$
$$
T_{\a\bh}{}^C = H_{\a\bh B}= \l^\a \l^\b R_{\a\gh\b}{}^\d=
 \hat\l^\ah \hat\l^\bh \hat R_{\g\ah\bh}{}^\dh= 0,$$
for any pure spinors $\l^\a$ and $\hl^\ah$ satisfying (\ref{pureII}). 
As in the heterotic case, the nilpotency constraints on $R_{ABC}{}^D$
are implied through Bianchi identities by the nilpotency constraints
on $T_{AB}{}^C$.

As will be discussed in section 4, the constraints of (\ref{nilII}) 
can be interpreted as Type II pure spinor integrability conditions and
imply
all the essential Type II supergravity constraints. The remaining conventional
Type II supergravity constraints will be implied by the holomorphicity
and antiholomorphicity of $\l^\a d_\a$ and $\hl^\ah \hd_\ah$.

\subsection{ Type II holomorphicity constraints }

To derive the constraints coming from holomorphicity of
$\l^\a d_\a$ and antiholomorphicity of $\hl^\a \hd_\a$,
first vary $\l^\a$, $w_\a$, $\hl^\ah$ and $\hat w_\ah$
in
(\ref{IIsm}) to obtain the equations
\be
\bar\p\l^\a = - (\Omega_{M\b}{}^\a \bar\p Z^M +
C_\b^{\a\gh} \hd_\gh + S_{\b\gh}^{\a\dh}\hl^\gh \hat w_\dh) \l^\b,
\label{dbarlII}
\ee
$$\bar\p w_\a =  (\Omega_{M\a}{}^\b \bar\p Z^M +
C_\a^{\b\gh} \hd_\gh + S_{\a\gh}^{\b\dh}\hl^\gh \hat w_\dh) w_\b, $$
$$\p\hl^\ah = - (\hat\Omega_{M\bh}{}^\ah \p Z^M +
\hat C_\bh^{\ah\g} d_\g + S_{\g\bh}^{\d\ah}\l^\g w_\d) \hl^\bh,$$
$$\p\bar w_\ah =  (\hat\Omega_{M\ah}{}^\bh \p Z^M +
\hat C_\ah^{\bh\g} d_\g + S_{\g\ah}^{\d\bh}\l^\g w_\d) \hat w_\bh.$$
And by
computing $E_\a^P (\d S/\d Z^P)$, 
one obtains the equation of motion
\be
\bar \p d_\a = E_\a^P [ (\p_{[P} E_{M]}^a E_N^b \eta_{ab}+ 
\p_{[P} E_{N]}^a E_M^b \eta_{ab} + \half H_{PMN}) \p Z^M \bar\p Z^N
\label{pbardII}
\ee
$$
+2 (\p_{[P} E_{N]}^\b d_\b + \p_{[P} \o_{N]\a}{}^{\b}
\l^\a w_\b)\bar\p Z^N 
+2 (\p_{[P} E_{N]}^\bh \hd_\bh +  \p_{[P} \hat\O_{N]\ah}{}^{\bh}
\hl^\ah \hat w_\bh )\p Z^N $$
$$
- \O_{P\a}{}^{\b}\bar\p (\l^\a w_\b) - 
\hat\O_{P\ah}{}^{\bh}\p (\hl^\ah \hat w_\bh) $$ 
$$+ \p_P P^{\a\bh} d_\a \hd_\bh +\p_P 
C_\a^{\b\gh}(Z) \l^\a w_\b \hd_\gh +
\p_P\hat C_{\ah}^{\bh\g}(Z) \hl^\ah \hat w_\bh d_\g +
\p_P S_{\a\gh}^{\b\dh}(Z) \l^\a w_\b \hl^\gh \hat w_\dh ]. $$

Putting these equations together, one finds
\be
\bar\p(\l^\a d_\a) = \l^\a[\half T_{\a Bc} (\Pi^B \bar \Pi^c + 
\Pi^c \bar\Pi^B)
+\half H_{\a B C} \Pi^B \bar\Pi^C
\label{putII}
\ee
$$
+ T_{\a B}{}^\b d_\b \bar\Pi^B + T_{\a B}{}^\bh \Pi^B \hd_\bh
+ (\nabla_\a P^{\rho\gh}+ C_\a^{\rho\gh} )d_\rho \hd_\gh 
+ R_{\a B \g}{}^\d \l^\g w_\d \bar\Pi^B
+ \hat R_{\a B \gh}{}^\dh \Pi^B \hl^\gh \hat w_\dh $$
$$+ \nabla_\a C_\b^{\d\gh} \l^\b w_\d \hd_\gh + 
\nabla_\a \hat C_{\bh}^{\gh\rho}
d_\rho \hl^\bh \hat w_\gh
+ \nabla_\a S_{\kappa\bh}^{\g\dh} \l^\kappa w_\g \hl^\bh \hat w_\dh +
(\nabla_\a \hat C_\bh^{\gh\rho}+
S_{\a\bh}^{\rho\gh})
d_\rho \hl^\gh \hat w_\bh], $$
where $\Pi^A = E^A_M \p Z^M$, $\bar\Pi^A = E^A_M \bar\p Z^M$,
$T_{ABc} = \eta_{cd}
T_{AB}{}^d$,
and all
superspace derivatives acting on unhatted spinor indices
are covariantized using the $\O_{M\a}{}^{\b}$ connection while
all superspace derivatives acting on hatted spinor indices
are covariantized using the $\hat\O_{M\ah}{}^\bh$ connection.
Furthermore, the torsion $T_{AB}{}^\a$ and curvature $R_{AB \g}{}^{\d}$
are defined as in (\ref{conn}) using the $\O_{M\a}{}^\b$ connection
whereas the torsion
$T_{AB}{}^\ah$ and curvature $\hat R_{AB\gh}{}^{\dh}$ are defined
using the $\hat\O_{M\ah}{}^\bh$ connection.
Note that $T_{Abc}$ appears only in the combination $T_{\a(bc)}$.
This combination is independent of
the spin connections 
since $\O_M^{(s)}$ and $\hO_M^{(s)}$
only act on spinor indices and since
$\O_M^{ab}$ and $\hO_M^{ab}$ are antisymmetric in their vector indices.

Plugging into (\ref{putII})
the equations of motion which come from
varying $d_\a$ and $\hd_\ah$,
\be
\bar\Pi^\a =  -P^{\a\bh}
\hat d_\bh- \hat C_\bh^{\gh\a}\hl^\bh \hat w_\gh, \quad
\Pi^\ah =  P^{\b\ah}
d_\b-  C_\b^{\g \ah} \l^\b w_\g,
\label{dbarZII}
\ee
one finds that holomorphicity of
$\l^\a d_\a$ implies that
\be
T_{\a (bc)} = H_{\a cd} =  H_{\a\bh\g} =
T_{\a\b c} + H_{\a\b c}=
T_{\a\bh c} - H_{\a\bh c}=0 
\label{plugII}
\ee
$$  T_{\a c}{}^\b + T_{\a\gh c} P^{\b\gh} =
 T_{\a c}{}^\bh - T_{\a\g c} P^{\g\bh} =
T_{\a\b}{}^\gh  -\half H_{\a \b \g} P^{\g\gh} = T_{\a\gh}{}^\b=0,$$
$$ C^{\g\bh}_\a + \nabla_\a P^{\g\bh} - T_{\a\rho}{}^\g P^{\rho\bh}=
\hat R_{c\a\bh}{}^\gh +T_{\a\rho c} \hat C_{\bh}^{\gh \rho}=
\hat R_{\a\d\bh}{}^\gh -\half H_{\a\d\rho} C_\bh^{\gh \rho} =0,$$
$$
S_{\a\gh}^{\rho\dh} + \hat R_{\a\bh\gh}{}^\dh P^{\rho\bh} +\nabla_\a 
\hat C^{\dh\rho}_\gh - T_{\a\b}{}^\rho \hat C_\gh^{\dh\b} =0, $$
$$ \l^\a\l^\b  (R_{c \a \b}{}^\g  + T_{\a\dh c}C_\b^{\g\dh})=
\l^\a\l^\b  R_{\dh \a \b}{}^\g = 0,$$
$$
\l^\a\l^\b  (\nabla_\a C^{ \d\gh}_\b -R_{ \a \k\b}{}^\d P^{\k\gh}) =
\l^\a\l^\b  (\nabla_\a S_{\b\gh}^{\rho\dh} - \hat 
R_{\a\hat\kappa\gh}{}^\dh C_\b^{\rho\hat\kappa} -
R_{\a\kappa\b}{}^\rho \hat C_\gh^{\dh\kappa})=0,$$
where the last two lines 
of equations must be satisfied for any pure spinor
$\l^\a$.
Antiholomorphicity of $\hl^\ah \hd_\ah$ implies the hatted
version of the above equations. The only subtle point is
that it implies $T_{\ah\bh c}-H_{\ah\bh c}= T_{\ah\b c} + H_{\ah\b c}=0$, 
which together
with the above equations implies that 
\be
T_{\a\b c}+ H_{\a\b c}= T_{\ah\bh c}- H_{\ah\bh c} = 
T_{\a\bh c} = H_{\a\bh c}=0.
\label{addhol}
\ee

The constraints of (\ref{nilII}) and (\ref{plugII}) will now
be shown to imply the correct Type II supergravity 
equations of motion.

\subsection{Type II supergravity constraints}   

The analysis of the Type II constraints of
 (\ref{nilII}) and (\ref{plugII}) will closely
resemble the analysis of the heterotic constraints in subsection
(2.5).
At scaling dimension
$-\half$, the constraints of (\ref{nilII}) imply
that 
\be
H_{\a\b\g}=H_{\a\b\gh}=H_{\a\bh\gh}=
H_{\ah\bh\gh}=0
\label{Hzero}
\ee
since there is no non-zero symmetric 
$H_{\a\b\g}$ and $H_{\ah\bh\gh}$ satisfying $\l^\a\l^\b H_{\a\b\g}=0$
and $\hl^\ah\hl^\bh H_{\ah\bh\gh}=0$. 

At dimension 0, the constraints $\l^\a\l^\b T_{\a\b}{}^c=\hl^\ah\hl^\bh
T_{\ah\bh}{}^c= 0$ imply that  
$T_{\a\b}{}^c = i(\g^d)_{\a\b} f_d^c$ and 
$T_{\ah\bh}{}^c =i (\g^d)_{\ah\bh} \hat f_d^c$ 
for
some $f_d^c$ and $\hat f^d_c$. 
Using the dimension zero H Bianchi identities and
the local Lorentz and scale transformations
of (\ref{localtwo}) for the unhatted and hatted spinor indices
independently, both $f_d^c$ and $\hat f_d^c$ can be gauge-fixed to $\d_d^c$.
After this gauge-fixing, the only remaining gauge invariance is a single local
Lorentz invariance which acts on all spinor and vector indices in the standard
fashion.
Combining with the other dimension 0 constraints of (\ref{nilII}) and
(\ref{plugII}), one has
\be
T_{\a\b}{}^c = - \eta^{cd} H_{\a\b d} =i(\g^c)_{\a\b}, \quad 
T_{\ah\bh}{}^c = \eta^{cd} H_{\ah\bh d} =i(\g^c)_{\ah\bh}, \quad
T_{\a\bh}{}^c=H_{\a\bh c}=0.
\label{scalarsII}
\ee

At dimension $\half$, the constraints $\l^\a\l^\b T_{\a\b}{}^\g=0$
and
$\hl^\ah\hl^\bh T_{\ah\bh}{}^\gh=0$
imply that $T_{\a\b}{}^\g
= f_c^\g (\g^c)_{\a\b}$ and
$T_{\ah\bh}{}^\gh
= \hat f_c^\gh (\g^c)_{\ah\bh}$ 
for some $f_c^\g$ and $\hat f_c^\gh$. 
Using the shift symmetries
of (\ref{wgauge}), both $f_c^\g$ and $\hat f_c^\gh$
can be gauge-fixed to zero so that
$T_{\a\b}{}^\g= T_{\ah\bh}{}^\gh =0$.
The other dimension $\half$ constraints,
\be
H_{\a c d}=T_{\a (cd)}=T_{\a \b}{}^\gh = T_{\a\bh}{}^\g= 0,
\label{otherhalf}
\ee
$$H_{\ah c d}=T_{\ah (cd)}=T_{\ah\bh}{}^\g= T_{\ah\b}{}^\gh =0,$$
imply through the Bianchi identities 
$(\nabla T + T T)_{\a\b\g}^c =0$
and
$(\nabla T + T T)_{\ah\bh\gh}^c =0$ that
\be
T_{\g b}{}^c= 2\eta^{cd}(\g_{bd})_\a{}^\b \O^{(s)}_\b,\quad
\hat T_{\gh b}{}^c= 2 \eta^{cd}(\g_{bd})_\ah{}^\bh \hO^{(s)}_\bh
\label{bianchit}
\ee
where $T_{\g b}{}^c$ is defined using the $\O_M^{bc}$ spin connection
and $\hat T_{\gh b}{}^c$ is defined using the $\hO_M^{bc}$ spin connection.
Furthermore, the Bianchi identities
$(\nabla T + T T)_{\a\bh\gh}^c =0$
and $(\nabla T + T T)_{\ah\b\g}^c =0$
imply that 
\be
\hat T_{\a b}{}^c= T_{\ah b}{}^c = \O_\ah^{(s)}= \hO_\a^{(s)} =0
\label{secbianch}
\ee
where
$\hat T_{\a b}{}^c$ is defined using the $\hO_M^{bc}$ spin connection
and $T_{\ah b}{}^c$ is defined using the $\O_M^{bc}$ spin connection.

At dimension one, the constraint $T_{c \a}{}^\b= T_{c\ah}{}^\bh=0$ 
decomposes into
\be
T_{c\a}{}^\b= T_c^{defg}(\g_{defg})_\a{}^\b+ T_c^{de}(\g_{de})_\a{}^\b
+ T_c \d_\a^\b=0,
\label{decompII}
\ee
$$T_{c\ah}{}^\bh= \hat T_c^{defg}(\g_{defg})_\ah{}^\bh+ \hat 
T_c^{de}(\g_{de})_\ah{}^\bh
+ \hat T_c \d_\ah^\bh=0.$$
The constraints
$T_c=\hat T_c=0$ and
$T_c^{de}=\hat T_c^{de}=0$ determine the vector components
of the spin connections $\O_c^{(s)}$, $\hO_c^{(s)}$,
$\O_c{}^{de}$ and $\hO_c{}^{de}$, whereas the
constraint
$T_c^{defg}=\hat T_c^{defg}=0$ is implied by the Bianchi identities
$(DH + TH)_{bc\alpha\gamma} (\g^{bdefg})^{\a\g}=0$
and
$(DH + TH)_{bc\ah\gh} (\g^{bdefg})^{\ah\gh}=0$.
The constraints $T_{\a c}{}^\bh = (\g_c)_{\a\g} P^{\g\bh}$ 
and
$T_{\ah c}{}^\b = (\g_c)_{\ah\gh} \hat P^{\b\gh}$ 
for some
$P^{\g\bh}$ and $\hat P^{\b\gh}$ are implied by the Bianchi identities
$(\nabla T +TT)_{\a\b\g}^\dh = 
(\nabla T +TT)_{\ah\bh\gh}^\d = 0$. And $P^{\g\bh}=
\hat P^{\g\bh}$ is implied by the Bianchi identity 
$(\nabla T + TT)_{\a\bh c}^c=0$.
Similarly, all other constraints in (\ref{nilII}) and (\ref{plugII}) are
either implied by Bianchi identities or define $C_\a^{\b\gh}$,
$\hat C_\ah^{\bh\g}$ and $S_{\b\dh}^{\a\gh}$ in terms of the supervielbein.

The above constraints imply that all background superfields appearing in
the action of (\ref{IIsm}) can be expressed in terms of the
spinor supervielbein $E_\a^M$ and $E_\ah^M$. Furthermore,
the constraints 
\be
T_{\a\b}{}^c=i(\g^c)_{\a\b},\quad T_{\ah\bh}{}^c= i (\g^c)_{\ah\bh},\quad
T_{\a\bh}{}^c=0
\label{scalarcons}
\ee
imply the on-shell equations of motion for $E_\a^M$ and $E_\ah^M$.
So the constraints of 
(\ref{nilII}) and (\ref{plugII}) imply the Type II supergravity
equations of motion. In the following section,
the above pure spinor description of Type IIB supergravity
will be related to the Howe-West (HW) description of \cite{hw}.
It should  similarly be possible to relate the pure spinor description
of Type IIA supergravity to the IIA superspace description of \cite{gatesIIA}.

\section{Relation with the $SL(2,\bbbold{R})$ 
Covariant Description of IIB Supergravity}

In this section we shall 
demonstrate that the 
constraints on the torsion for IIB derived in the 
preceeding section are indeed equivalent to the 
HW equations of motion of IIB supergravity described in \cite{hw}. 
We shall do this by first showing that the latter 
are generated by the standard dimension 
zero torsion constraint and then 
exhibiting the explicit transformation from the 
standard IIB superspace torsions to those derived 
above. In order to carry through the first step we use the 
method of Weyl superspace and then we reduce the structure 
group to the Lorentz group. In order to establish the result 
fully we also have to examine the scalars in the theory.

The complete IIB supergravity theory was derived from a superspace
perspective in \cite{hw}. However, although complete results were
given there for all of the superspace field strength tensors, no
attempt was made to identify a minimal generating set of
constraints. Moreover, the HW formalism is manifestly
locally $U(1)$ and globally $SL(2,\bbR)$ invariant and this is not
convenient for the applications we have in mind here. We shall work
initially in an $SO(2)$ formalism (rather than $U(1)$) since this
will be easier to adapt to our purposes.

For IIB superspace we use the
same HW conventions as in \cite{hw}, although we use $\c$ to denote the
$16\xz 16$ spin matrices instead of $\s$. To convert $SO(2)$ spinor indices
$i,j,...$
to
$U(1)$ indices, we write
\be
 v^i\rightarrow v^{\pm}={1\over\sqrt2}(v^1 \pm v^2)
\ee
and
\be
 v_i\rightarrow v_{\pm}={1\over\sqrt2}(v^1 \mp v^2).
\ee
So the metric and $\e$-tensor are
\be
 \d_{+-}=1,\qquad \e^{+-}=-i,\qquad \e_{+-}=i,\qquad \e_+{}^+=i.
\ee
The summation is therefore $u^i v_i= u^+ v_+ + u^- v_-$.
\footnote{This causes a slight problem in the superspace summation
convention which should be taken to be $u^{\a i} v_{\a i}=u^{\a +}
v_{\a +} + u^{\a -}v_{\a -}$, whereas in \cite{hw} one finds $u^{\a}
v_{\a}- u^{\bar\a} v_{\bar\a}$. So, in converting from HW conventions to
$SO(2)$, one has to remember to insert an extra minus sign for
downstairs $\a -$ indices.
This means, for example, that we must take
$ T_{\a i\b j}{}^c= i\d_{ij}(\c^c)_{\a\b}$
since then one finds
$ T_{\a +\b -}{}^c=i(\c^c)_{\a\b}\Rightarrow T_{\a\bar\b}{}^c=-i(\c^c)_{\a\b}$
in agreement with \cite{hw}.}
To convert 
$SO(2)$ vector indices $r,s,...$ to $U(1)$ indices, we have
\be
 v_r=(\tau_r)^{ij} v_{ij} \leftrightarrow v_{ij}=(\tau^r)_{ij} v_r
\ee
where
\be
 \tau^r ={1\over\sqrt2}(\s_3,\s_1).
\ee
We can then put
\be
 v^{\pm\pm}={1\over\sqrt2}(v^1 \pm v^2)
\ee
for vector indices and this is consistent since $(\tau^{++})_{++}=1$.

In subsection (4.1), we shall first show that
the equations of motion of IIB supergravity follow (up to
topological niceties) from the usual dimension zero constraint
 \be
 T_{\a i\b j}{}^c= i\d_{ij}(\c^c)_{\a\b}.
\label{dimz}
 \ee
We shall do this by working in Weyl superspace, i.e. we shall
include a scale factor in the structure group. Following through
the consequences of this we find that the scale curvature vanishes
so that the scale connection is pure gauge. If we then take it to
vanish we recover the equations of \cite{hw}. This procedure is very
similar to the approach used in \cite{h2} to prove that the
equations of motion of $D=11$ supergravity follow from the standard
dimension zero constraint.

Since the standard dimension zero constraint of (\ref{dimz})
 is required by the
nilpotency of $Q$, it then follows that the equations of motion of
IIB supergravity are indeed implied in the pure spinor formalism. 
However, as we have
seen, there are many other equations at dimensions greater than zero
that are required to hold either by the nilpotency or by the
holomorphicity of $Q$. In subsection (4.2),
we shall check these explicitly at
dimension one-half by comparing our results with those of section 3.

In the HW superspace description of Type IIB supergravity,
$SL(2,\bbR)$ global symmetry is manifest since the two scalars are
described by an $SO(2)\bsh SL(2,\bbR)$ coset. However, this
$SL(2,\bbR)$ symmetry is not manifest in the pure
spinor description since the dilaton and axion do not appear in
an $SL(2,\bbR)$ covariant manner. In subsection (4.3), we
will relate these two descriptions of the Type IIB scalars and will
show that the target-space metric appearing in the pure
spinor version of the Type IIB sigma model is in
string gauge.

\subsection{Weyl superspace}

To get the superspace constraints under control it is useful to
include a scale factor in the connection. The structure group is
then $Spin(1,9)\times Spin(2)\times \bbR^+$. The full connection
(denoted by a tilde) is
\bea
 \tO_{a}{}^b&=&\O_{a}{}^b+ 2\d_a{}^b \P \\
 \tO_{\a i}{}^{\b j}&=& \d_i{}^j\O_{\a}{}^{\b} +
 \d_{\a}{}^{\b}(\d_i{}^j\P + \e_i{}^j \S)
\label{Uonec}
\eea
where $\O,\S,\P$ are respectively the connections for the Lorentz,
$U(1)(=Spin(2))$ and scale factors. We shall use the notation $\O'$
to denote the $Spin(1,9)\xz U(1)$ connection, so $\O'\sim \O + \S$.
Similarly, for the curvatures we have $\tR\sim R'+M\sim R + M + N$
where $M$ and $N$ are respectively the $U(1)$ and scale curvatures.

At dimension one-half we find, using the Bianchi identity
 \be
 \tD_{(\a i} \tT_{\b j\c k)}{}^d + \tT_{(\a i\b j}{}^E \tT_{|E|\c k)}{}^d=0
 \ee
and
the freedom to choose the dimension
one-half components of the connection and 
the even basis vectors $E_a$, that the dimension one-half
component of the torsion tensor is
\be
 \tT_{\a i\b j}{}^{\c k}=-i(\c^a\c_a -2\d\d)_{(\a\b)}{}^{\c\d}\L_{\d ij}{}^k
\label{Ldef}
\ee
where $\L_{\a ijk}$ is totally symmetric and traceless on its
$Spin(2)$ indices, while
\be
\tT_{\a b}{}^c=0.
\ee
This is exactly the same as in \cite{hw}, and we
identify the HW spinor field $\L$ by
\bea
 \L=\sqrt2(\L_{222}+i\L_{111}) &=& i\L_{---} \\
 \bar\L=\sqrt2(\L_{222}-i\L_{111}) &=& -i\L_{+++}.
\eea

At dimension one one has to solve two Bianchi identities
 \bea
 \tR_{\a i\b j,c}{}^d&=&\tT_{\a i\b j}{}^E \tT_{E c}{}^d + \tT_{c\a
 i}{}^{\e m} \tT_{\e m \b j}{}^d + \tT_{c\b
 j}{}^{\e m} \tT_{\e m \a i}{}^d \nn\\
 \tR_{(\a i \b j,\c k)}{}^{\d l}&=&\tD_{(\a i} \tT_{\b j \c k)}{}^{\d l}
 +\tT_{(\a i\b j}{}^e \tT_{e \c k)}{}^{\d l}+
 \tT_{(\a i\b j}{}^{\e m} \tT_{|\e m|\c k)}{}^{\d l}.
\eea
After a long and tedious calculation one can show that the only
non-zero dimension one components of the curvature and torsion
tensors are those which correspond to the dimension one components
of the IIB supergravity multiplet, that is $F_{abc}, P_a,
G_{abcde}$ together with fermion bilinear terms. The tensors $F,P$
and $G$ are asociated with the antisymmetric tensor gauge fields of
the theory and the scalar fields ($P_a$ is essentially the
derivative of the scalar fields). One also determines the spinorial
derivative of $\L$ and the dimension one component of the $U(1)$
curvature $M$ in terms of these physical fields. Moreover, one
finds that the dimension one component of the scale curvature $N$
vanishes, $N_{\a i\b j}=0$. From this, one immediately concludes
with the aid of the scale curvature Bianchi identity, $dN=0$, that
the whole of $N$ vanishes and so the scale connection is pure gauge
as anticipated. At this stage we can set the scale connection equal
to zero and recover the HW torsions 
and curvatures of \cite{hw}. From these results 
one can then construct super extensions of 
$F,G,P$ which satisfy corresponding Bianchi 
identities. In particular, one can deduce the 
existence of the two scalar fields 
described by an $SL(2,\bbR)\bsh U(1)$ coset space.

\subsection{Lorentz superspace}

To recover the form of the torsion and curvature tensors derived
from the pure spinor formalism,
we need firstly to restrict the structure group to be the
ten-dimensional spin group. This means that the components of $\P$
and $\S$ will appear in the redefined torsion. 
Moreover, we shall choose a different scale gauge from
$\P=0$ which means that $\P=-dS$ for some 
scalar field $S$ and also that there is change of 
basis with respect to the HW basis, i.e. 
$E^a=e^{2s} E^a_{HW}$, etc. Explicitly, we have
 \be
 \tT_{AB}{}^C = T_{AB}{}^C + 2\P_{[A} I_{B]}{}^C +2\S_{[A}
 J_{B]}{}^C
 \ee
where
 \bea
 I_A{}^B&=&\cases{ I_a{}^b=2\d_a{}^b\cr I_{\a i}{}^{\b
 j}=\d_{\a}{}^{\b}\d_i{}^j} \nn\\
 J_A{}^B&=&\cases{J_a{}^b=0\cr J_{\a
 i}{}^{\b j}=\d_{\a}{}^{\b}\e_i{}^j}
 \eea
and where the mixed spinor-vector components of $I$ and $J$ are
zero. In particular, at dimension one-half, we have
\bea
\tT_{\a i\b j}{}^{\c k}&=&T_{\a i\b j}{}^{\c k} 
+\d_{\a}{}^{\c}\left(\d_i{}^k\P_{\b j} + \e_i{}^k \S_{\b j}\right)
+\d_{\b}{}^{\c}\left(\d_j{}^k\P_{\a i} + \e_j{}^k \S_{\a i}\right)\\
\tT_{\a i b}{}^c&=&T_{\a i b}{}^c + 2\d_b{}^c \P_{\a i}.
\eea

We shall also have to shift the Lorentz connection as 
\be
\O_{\a bc}\rightarrow 
\O^{(1,2)}_{\a bc}=\O_{\a bc}+(\c_{bc})_{\a}{}^{\b}Y^{(1,2)}_{\b}.
\ee
The notation here is that the connection labelled $i=1,2$ will 
act on spinor indices with the same internal index label. 
Since the two connections will be different, this procedure 
manifestly breaks $SO(2)$. For the moment we shall suppose 
that the vector indices are acted upon by the original $\O$. 
Finally, in order to make a direct comparison to the 
earlier results we shall have to shift the vectorial basis $E_a$ by
\be
E_a\rightarrow E_a + i(\c^a)^{\a\b}\chi_{\a i} E_{\b i}.
\ee

If we choose
\bea
Y^{(1)}_{\a 1} &=& i\L_{\a 111}\nn, \\
Y^{(1)}_{\a 2}&=&-i\L_{\a 222}\nn, \\
Y^{(2)}_{\a 1} &=& -i\L_{\a 111}\nn, \\
Y^{(2)}_{\a 2} &=& i\L_{\a 222}\nn,\\
\chi_{\a 1}&=& -i\L_{\a 111}\nn, \\
\chi_{\a 2}&=& -i\L_{\a 222},
\eea
and if, in addition,
\bea
\S_{\a 1}&=&-i\L_{\a 222}\nn,\\
\S_{\a 2}&=&i\L_{\a 111}\nn,\\
\P_{\a 1}&=&-{i\over2}\L_{\a 111}\nn,\\
\P_{\a 2}&=&-{i\over2}\L_{\a 222},
\la{SP}
\eea
then we find that all components of 
the redefined $T_{\a i\b j}{}^{\c k}$ vanish except for
\bea
T_{\a 1\b 1}{}^{\c 1}&=&-2\d_{(\a}{}^{\c}\O^{(s)}_{\b)}\nn , \\
T_{\a 2\b 2}{}^{\c 2}&=&-2\d_{(\a}{}^{\c}\hat\O^{(s)}_{\b)}
\label{spinvanish}
\eea
where
\be
\O^{(s)}_{\a}=i\L_{\a 111};\qquad \hat\O^{(s)}_{\a}=i\L_{\a 222}.
\ee
If we also define new vectorial 
torsions with respect to the new connections $\O^{(1,2)}_{\a bc}$, we find
\bea
T^{(1)}_{\a 1 b}{}^c&=& 2(\c_b{}^c \O^{(s)})_{\a}\nn , \\
T^{(1)}_{\a 2 b}{}^c&=& 0\nn,\\
T^{(2)}_{\a 1 b}{}^c&=& 0\nn,\\
T^{(2)}_{\a 2 b}{}^c&=& 2(\c_b{}^c \hat\O^{(s)})_{\a}.
\label{vectorvan}
\eea

We shall verify that equations \eq{SP} are indeed satisfied 
in the next section. For the moment, assuming that they are, 
we are now in a position to compare directly with the 
dimension one-half results coming from the pure
spinor formalism. In order to do this, 
we remove the fermionic scale connection from the type II 
structure group. After identifying
the indices $(\a 1,\a 2)=(\a,\ah)$, we 
find that the only non-vanishing components 
of the redefined torsion with three spinorial indices are
those of (\ref{spinvanish}).
The vectorial torsions, which do not need to be redefined, are
those of (\ref{vectorvan}). 
% \bea
% T_{\a  b}{}^c&=& 2(\c_b{}^c \O^{(s)})_{\a}\nn \\
% T_{\ah b}{}^c&=& 0\nn\\
% \hT_{\a  b}{}^c&=& 0\nn\\
% \hT_{\ah b}{}^c&=& 2(\c_b{}^c \hat\O^{(s)})_{\ah}
% \eea
%
%The two sets of results are clearly in agreement if we 
%identify the indices $(\a 1,\a 2)=(\a,\ah)$ and 
%take $\O^{(s)}=\psi$, $\hat\O^{(s)}=\hat\psi$.
We have therefore succeeded in demonstrating that the torsions 
derived from the pure spinor formalism are 
indeed in agreement with those of \cite{hw}
after suitable field redefinitions.
To complete the picture we must verify that $\P=-dS$ 
and that the expressions given for $\S_{\a i}$ in \eq{SP} are 
correct. To do this, we need to examine the scalar fields in the theory.

\subsection{Scalar fields}

The scalar fields take their values in the space $SO(2)\bsh
SL(2,\bbR)$. We describe then by a real two by two matrix $\cU$
acted on by $\cU\rightarrow h\cU g^{-1}$, for $h\in SO(2),\ g\in
SL(2,\bbR)$. In index notation we write $\cU_r{}^R$. Note that $r$
is vector $SO(2)$ index while $R$ is an $SL(2,\bbR)$ doublet index. The
Maurer-Cartan form $\cM$ is given by
 \be
 \cM= d\cU\,\cU^{-1}.
 \la{s6}
 \ee
Since it is Lie-algebra valued it can be written as
 \be
 \cM_r{}^s=P_r{}^s + 2 \e_r{}^s \S
 \la{s7}
 \ee
where $\Sigma$ is the $U(1)$ connection of (\ref{Uonec}) and
$P^{rs}$ is symmetric traceless, i.e. in complex notation we have
$P^{++++}:=P$ where $P$ is the HW one-form defined in \cite{hw}. The
Maurer-Cartan equation, $d\cM + \cM^2=0$, implies that
 \bea
 DP&=& 0 \\
 M&=& -{i\over2} P\wedge\bar P
 \la{s8}
 \eea
where $M$ is the $U(1)$ curvature tensor.

There is a HW $SL(2\bbR)$ doublet of three-form field strength
tensors $\tF_R$, and we define $F:=\cU \tF$. Assuming that
$d\tF=0$ we find that
 \be
 D F_r = P_r{}^s F_s
 \la{s9}
 \ee
As before, we can identify $F_{--}=F^{++}$ with the field $F$ of
\cite{hw}. In a complex basis \eq{s9} reads
 \be
 DF = P\wedge\bar F
 \la{s10}
 \ee
as in \cite{hw}.

The field $\cU$ is not quite the same as the HW field $\cV$.
The two are related by
 \be
 \cU = \e \cV^{-1} \e.
 \la{s11}
 \ee
The Maurer-Cartan form is then
 \be
 \ba{ccl}
 d\cU\, \cU^{-1}&=&\e \cV^{-1} d\cV \e \\
 &&\\
 &=& \left(\ba{rr}0 & 1\\ -1 & 0\ea\right)
     \left(\ba{rr} 2i\S & P\\ \bar P & -2i\S\ea\right)
     \left(\ba{rr}0& 1\\ -1 & 0\ea\right)\\
 &&\\
 &=& \left(\ba{rr}2i\S & \bar P \\ P & -2i\S\ea\right).
 \ea
 \la{s12}
 \ee
In the second line of (\ref{s12}),
we have used the formula for the Maurer-Cartan
form in \cite{hw}
(with $\S$ instead of $Q$), and in the final line we have the
correct expression in the new conventions in a complex basis.

In the physical gauge we can write the components of $\cU$ in terms
of $\tau:=\tau_1 +i\tau_2:=C_0 + i e^{-\Phi}$ where $\Phi$ is the dilaton and
$C_o$ is the axion. In the real basis we have
 \be
 \cU={1\over \sqrt{\tau_2}}\left(\ba{rr} 1 & \tau_1 \\ 0& \tau_2\ea\right)
 \la{s13}
 \ee
and one can check that $\tau$ has the expected transformation under
$SL(2,\bbR)$, i.e.
 \be
 \tau\rightarrow {a\tau + b\over c\tau + d}
 \la{s14}
 \ee
where $g\in SL(2,\bbR)$ is
 \be
 g=\left(\ba{rr} a & -b\\ -c & d\ea\right).
 \la{s15}
 \ee
In a complex basis (for both indices)
 \be
 \cU={1\over 2\sqrt{\tau_2}}\left(\ba{rr} 1+i\bar \tau & 1-i\bar \tau \\
 1+i\tau & 1-i\tau \ea\right).
 \la{s16}
 \ee

If one computes the Maurer-Cartan form  in this gauge one finds
 \bea
 P&=&{1\over2}(d\Phi +i e^{\Phi} d C_o) = {id\tau\over 2\tau_2},\\
 \S&=& {1\over4} e^{\Phi} d C_o.
 \la{s17}
 \eea
In the HW description, $P^{++++}$ is related to
$\L_\a^{jkl}$ of (\ref{Ldef}) by 
 \be
 P^{++++}=2 E^{\a +} \L^{+++}_{\a} + E^a P_a^{++++}
 \la{s18}
 \ee
 and $\tau$
is chiral, i.e. $\bar D\tau=0$. This implies
 \bea
 D_{\a 1} C_o &=& -e^{-\Phi} D_{\a 2}\, \Phi \\
 D_{\a 2} C_o &=& e^{-\Phi} D_{\a 1}\, \Phi
 \la{s19}
 \eea

Using this we can express the components of $\L$ as
 \bea
 \L_{\a 111}&=& -{i\over4} D_{\a 1} \Phi, \\
 \L_{\a 222}&=& -{i\over4} D_{\a 2} \Phi.
 \la{s20}
 \eea
We can also express the components of $\S$ in terms
of $D\Phi$ as
 \bea
 \S_{\a 1}&=& -{1\over4} D_{\a 2}\Phi ,\\
 \S_{\a 2} &=& {1\over4} D_{\a 1}\Phi.
 \la{s21}
 \eea

Now earlier we found what the $U(1)$ and scale connections had to
be chosen to be at dimension one-half in order to achieve
$Q$-integrability. We required
 \bea
 \S'_{\a 1} &=& -i\L'_{\a 222}, \\
 \S'_{\a 2} &=& i\L'_{\a 111} , \\
 \P'_{\a 1} &=& {-i\over2}\L'_{\a 111} , \\
 \P'_{\a 2} &=& {-i\over2}\L'_{\a 222},
 \la{s22}
 \eea
where $\P$ is the scale connection and where the prime indicates
the basis which is related to the unprimed HW one by $E'^{\a i}= e^S
E^{\a i},\ E'^a= e^{2S}E^a$. We also required $\P=-dS$, since it is
pure gauge. So we can identify
 \be
 S={\Phi\over 8}.
 \la{s23}
 \ee

In addition, if we compare the expressions for the components of
$\L$ and $\S$ in terms of $D\Phi$, we see that they agree, and so
everything works as expected. If we define bosonic metrics by
 \bea
 G&=& E^b\otimes E^ a \h_{ab}, \\
 G'&=& E'^b\otimes E'^ a \h_{ab},
 \la{s24}
 \eea
then $G'=e^{{\Phi\over2}}G$. This means we can identify $G'$ with the
string metric and $G$ with the Einstein metric, so the conformal
transformation we need to make is precisely the one which goes
between the two frames.

\section{Higher Order $\a'$ Corrections}

In this paper we have verified to lowest order in $\a'$ that nilpotence
and holomorphicity of the pure spinor BRST operator implies the superspace
equations of motion for the background supergravity fields. The
next question to investigate is how these superspace equations of motion
are modified by higher order $\a'$ corrections to the nilpotency
and holomorphicity conditions. Since the sigma model is a free action
in a flat background, one can compute these corrections using standard
sigma model methods by separating
the worldsheet variables into classical and quantum parts and expanding
in normal coordinates around a flat background.

When the background fields satisfy their string-corrected equations of
motion,
one expects that the $\b$-functions
of the sigma model should vanish, i.e. that the sigma model remains
conformally invariant at the quantum level. However, unlike the bosonic
string sigma model, quantum conformal invariance is not expected to
imply the complete set of equations of motion for the background fields.
In addition, one needs to impose the conditions that,
at the conformal fixed point, 
$\l^\a d_\a$ is holomorphic and nilpotent. It should be possible
to impose these nilpotence and holomorphicity
conditions perturbatively in $\a'$ by computing
contributions of the quantum worldsheet variables and the Fradkin-Tseytlin
term to the equations of motion and OPE's of $\l^\a$ and $d_\a$.

The necessity of imposing BRST nilpotence and holomorphicity 
can be seen at the linearized level by analyzing the superstring
vertex operators of (\ref{vsg}), (\ref{vsym}) and (\ref{vsgII}).
When the superfields in these vertex operators are on-shell to linearized
order, one can check that the vertex operators 
have no poles with the stress tensor $T$ and therefore preserve quantum
conformal invariance. However, the condition of having no poles
with $T$ is weaker than BRST invariance (i.e. $[Q_{flat}, V]=0$)
and does not imply the complete set of linearized on-shell conditions.
Note that $Q= Q_{flat}+V$ to linearized level, so $[Q_{flat},V]=0$
implies that $Q$ is nilpotent to linearized order.

Besides the Chern-Simons
modifications to the three-form field
strength, 
the first superstring corrections to the supergravity equations
of motion are expected to come at order $(\a')^3$, e.g. from the 
$R^4$ term. Since the supergravity equations of motion
are
implied by classical
nilpotence and holomorphicity of the BRST operator, one expects to see
these $(\a')^3$ corrections to the equations at
three loops in the nilpotence and holomorphicity conditions.
However, already at first order in $\a'$, there are several non-trivial
one-loop contributions to the nilpotence and holomorphicity conditions which
must be cancelled by contributions from the Fradkin-Tseytlin term and
from the Chern-Simons modification to the three-form field strength.

For example, for the heterotic superstring, the term 
$E_\a^P(- \O_{P \g}{}^\b \bar\p(\l^\g w_\b)-A_{P I}\p\bar J^I)$ appearing in
$\bar\p d_\a$ in equation (\ref{barpd}) gets one-loop corrections from
the chiral anomalies
\be
\p \bar J^I= \half\a' \p_{[M} A_{N]}^I \p Z^M \bar \p Z^N, 
\label{chanom}
\ee
\be
\bar\p (\l^\g w_\b) =-\half \a' \p_{[M} \O_{N]\b}{}^\g \p Z^M\bar \p Z^N
+ {1\over 8}\a' r \d_\b^\g
\label{hetanom}
\ee
where $r$ is the worldsheet curvature and the coefficient ${1\over 8}\a'$
in (\ref{hetanom})
can be obtained by computing the coefficient of the
triple pole of $\l^\g w_\b$
with the pure spinor stress tensor $T_\l$ and dividing by 
four.\footnote{The triple pole of
$\l^\g w_\b$ with $T_\l$ can be computed using the formulas
of \cite{relating} where
$\l^\a w_\a= 2\a' \p h$ is the ghost-number current,
$T_\l= {1\over {10}} N_{ab} N^{ab} -\half (\p h)^2 -2\p^2 h$ is the
pure spinor stress tensor, and $h(y)h(z)\to -\log(y-z)$.}
So $\bar\p d_\a$ gets a 
one-loop contribution 
\be
-2 \a' r E_\a^P \Omega_P^{(s)} +\half\a' 
E_\a^P(\O_{P\g}{}^\b \p_{[M} \O_{N]\b}{}^\g
- A_P^I \p_{[M} A_{N]}^I) \p Z^M \bar\p Z^N.
\label{hetloop}
\ee
After including other one-loop contributions coming from contractions
of the quantum worldsheet variables, one expects that the second term
in (\ref{hetloop}) is completed to 
$\half \a' E_\a^P w^{(CS)}_{PMN} \p Z^M\bar\p Z^N$ where
\be
w^{(CS)}_{PMN} = 3~ Tr (\O_{[P} \p_M \O_{N]} 
+{2\over 3} \O_{[P}\O_M\O_{N]} - 
 A_{[P} \p_M A_{N]}
-{2\over 3}  A_{[P} A_M A_{N]})
\label{CSform}
\ee
is the Chern-Simons three-form constructed
from the gauge, scale and Lorentz connections.

The first term in (\ref{hetloop})
is cancelled by the contribution from the 
Fradkin-Tseytlin term 
$${1\over{2\pi\a'}}\int d^2 z ~\half \a' r \Phi(Z),$$
 which contributes
$\half\a' r D_\a\Phi$ to $\bar\p d_\a$. So the $\a' r$ contribution
to $\bar\p d_\a$ is cancelled if the heterotic dilaton
superfield $\Phi$ is related to the scale connection $\O_P^{(s)}$ by
\be
D_\a\Phi= 4 E_\a^P\O_P^{(s)},
\label{phihet}
\ee
which can be checked to imply that the metric is in string gauge.

Since $\bar\p d_\a$ of (\ref{barpd}) also
contains the term $\half E_\a^P H_{PMN}
\p Z^M\bar \p Z^N$, the
second term in (\ref{hetloop}) can be cancelled
by redefining  
\be
H_{PMN}\to H_{PMN} - \a' w^{(CS)}_{PMN}.
\label{csterm}
\ee
As in the RNS sigma model \cite{HullWitten}, 
the need for redefining $H_{PMN}$ can
also be seen by requiring gauge invariance of the sigma model action.
Because of (\ref{chanom}) and (\ref{hetanom}), the action of 
(\ref{hetsm}) is invariant under local gauge, scale and Lorentz transformations 
only if
$B_{MN}$ is defined to transform as 
\be
\d B_{MN} = \a' (\p_{[M} A_{N]}^I \L^I - \p_{[M}\O_{N]}^{(s)} \L^{(s)} -
\p_{[M}\O_{N]}^{ab} \L_{ab})
\label{gaugeB}
\ee
where $\L^I$, $\L^{(s)}$ and $\L^{ab}$ are the gauge parameters. 

Similarly, for the Type II superstring, the anomalies 
\be
\bar\p (\l^\g w_\b) =-\half \a' \p_{[M} \O_{N]\b}{}^\g \p Z^M\bar \p Z^N
+ {1\over 8}\a' r \d_\g^\b,
\label{gnamtwo}
\ee
$$\p (\hl^\gh \hat w_\bh) =\half \a' \p_{[M} \hat\O_{N]\bh}{}^\gh 
\p Z^M\bar \p Z^N
+ {1\over 8}\a' r \d_\gh^\bh$$
imply from equation
(\ref{pbardII}) that the Type II dilaton superfield $\Phi$
is related to the scale connections $\O_P^{(s)}$ and
$\hat\O_P^{(s)}$ by
\be
D_\a\Phi= 4 E_\a^P\O_P^{(s)}, \quad
D_\ah\Phi= 4 E_\ah^P\hat\O_P^{(s)}.
\label{phiII}
\ee
One can check for the Type IIB superstring
that (\ref{phiII}) agrees with the relation found
in equation (\ref{s20}), which confirms that the metric is in string gauge.
Furthermore, the terms in (\ref{gnamtwo}) suggest that one should
redefine the Type II three-form field strength as
\be
H_{PMN}\to H_{PMN} - \a' (w^{(CS)}_{PNM} -\hat w^{(CS)}_{PNM})
\label{csIIterm}
\ee
where $w^{(CS)}$ is a Chern-Simons three-form constructed from
the unhatted spin connections $\O_P^{(s)}$ and $\O_P^{ab}$, and
$\hat w^{(CS)}$ is a Chern-Simons three-form constructed from
the hatted spin connections $\hat\O_P^{(s)}$ and $\hat\O_P^{ab}$.
However, since the differences of the vector components of the
spin connections, $\Omega_c^{(s)} - \hat \Omega_c^{(s)}$
and $\Omega_c^{ab} - \hat \Omega_c^{ab}$, are expected to vanish on-shell,
the vector components of the three-form, $H_{abc}$, are not expected
to be affected by (\ref{csIIterm}).

It would be interesting to verify that these and other
one-loop corrections to the BRST nilpotency and
holomorphicity conditions are cancelled by
the Fradkin-Tseytlin term and the Chern-Simons
modifications to the three-form. It would also be interesting
to verify that 
the sigma model actions of (\ref{hetsm})
and (\ref{IIsm}) are indeed conformally invariant at the quantum level
when the background fields are on-shell.

\vskip 30pt

{\bf Acknowledgements:}

NB would like to thank CNPq grant 300256/94-9,
Pronex grant 66.2002/1998-9, FAPESP grant 99/12763-0, and the
Clay Mathematics Institute for partial financial support. 
The research of PSH was supported in part by PPARC SPG grant 613.

\vfill\eject

 \end{document}